\documentclass[amssymb,aps,pre,preprint,secnumarabic]{revtex4}
\usepackage{amsmath}
\usepackage{amssymb}
\usepackage{graphicx}

\begin{document}

\title{Fluctuating Noise Drives Brownian Transport}

\author{Yoshihiko Hasegawa}

\affiliation{Department of Biophysics and Biochemistry, Graduate School of Science,
The University of Tokyo, Tokyo 113-0033, Japan}

\author{Masanori Arita}

\affiliation{Department of Biophysics and Biochemistry, Graduate School of Science,
The University of Tokyo, Tokyo 113-0033, Japan}

\date{Sep 13, 2012}
\begin{abstract}
The transport properties of Brownian ratchet was studied in the presence
of stochastic intensity noise (SIN) in both overdamped and underdamped
regimes. In the overdamped case, analytical solution using the matrix
continued fraction method revealed the existence of a maximum current
when the noise intensity fluctuates on intermediate time scale regions.
Similar effects were observed for the underdamped case by Monte Carlo
simulations. The optimal time-correlation for the Brownian transport
coincided with the experimentally observed time-correlation of the
extrinsic noise in \emph{Esherichia coli} gene expression and implied
the importance of environmental noise for molecular mechanisms. 
\end{abstract}
\maketitle

\section{Introduction}

Noise-induced phenomena are attracting much attention not only in
engineering but also in molecular biology. Counter-intuitively, noise
can enhance system performance by increasing transmission and synchronization
of information through stochastic resonance \cite{Benzi:1981:SR,McNamara:1989:SR,Jung:1991:AmpSR,Gammaitoni:1998:SR,McDonnell:2008:SRBook,McDonnell:2009:SR}
and noise-induced synchronization \cite{Marchesoni:1996:SpatiotempSR,Teramae:2004:NoiseIndSync,Acebron:2005:KuramotoReview,Nakao:2007:NISinLC}.

For example, molecular systems function efficiently in nano-scale
environments under multi-scale noise through thermal and other environmental
fluctuations \cite{Koern:2005:GeneNoiseReview,Patnaik:2006:NoiseReview,Maheshri:2007:LivingNoisyGene,Rausenberger:2009:GeneNoiseReview}.
This efficiency should not be interpreted by assuming steady-state
or Gaussian distribution. A recent single-cell observation of \emph{Escherichia
coli} revealed that the protein copy number does not obey the gamma
distribution~\cite{Taniguchi:2010:SingleCell}, and its stationary
distribution can be approximated by superstatistics (i.e., the superposition
of multiple statistical models; see Equation~\ref{eq:superstat_def}
and Discussion). Chabot \emph{et al.}\@ revealed that the cellular
variability in gene expressions comes from temporal (periodic) noise
which is related to circadian oscillation~\cite{Jeffrey:2007:CyanoBac}.
In accordance with such experimental observations, theoretical studies
also conclude that biochemical noise is not Gaussian to facilitate
enhanced functionality \cite{Nozaki:1999:ColoredNoiseNeuron,Fuentes:2001:nonGaussSR}.
Both experimental and theoretical approaches suggest fundamental roles
of noise-enhanced phenomena to render efficient molecular systems.

In this paper, we investigate the efficiency of \emph{Brownian motor}~\cite{Magnasco:1993:ThermalRatchet,Hanggi:2009:BrownianMotorsReview}
(or a ratchet transport) under a noisy, nonequilibrium state. It is
known that the violation of detailed balance induces a transport effect,
which is a model for many biological mechanisms including ion pumps
\cite{Astumian:1994:MolecularMotor,Astumian:1997:BrownMotor,Astumian:1998:MotorAndPump}
that use ATP for transport (In theory, however, the conformational
fluctuation of such pumps can facilitate transport even without ATP
\cite{Tsong:2002:NaK_ATPase}). 

Brownian transport has been intensively studied including in mass
separation \cite{Marchesoni:1998:MassSep}, electron transport in
a quantum ratchet \cite{Linke:2002:QuantumRatchet}, transport of
atoms in optical traps \cite{Lundh:2005:RatchetOptLat}, a random
walker model \cite{Jose:2005:RandomWalker}, and non-Gaussian noise
models \cite{Bouzat:2004:nonGaussMotor,Mangioni:2008:WalkerRatchetNGN}
(for more details, see comprehensive reviews \cite{Frey:2005:BrownianRev,Hanggi:2009:BrownianMotorsReview}).
One of the most popular ratchet models is the correlation ratchet
in which Brownian particles in a ratchet potential are driven by the
\textbf{addition} of white and colored noise. The model studied here
is a ratchet driven by the \textbf{multiplication} of white and colored
noise. Let us introduce our model more formally.

Superstatistics with temporal and/or spatial fluctuations \cite{Wilk:2000:NEXTParam,Beck:2001:DynamicalNEXT,Beck:2003:Superstatistics,Beck:2011:Superstatistics}
is used to explain non-Gaussian distributions \cite{Tsallis:1988:Generalization,Tsallis:2009:NonextensiveBook}
in applied physics. This concept is also seen in stochastic processes
in which noise fluctuation is treated in a static way~\cite{Beck:2001:DynamicalNEXT,Queiros:2005:VolatilityNEXT,Beck:2006:SS_Brownian,Rodriguez:2007:SS_Brownian,Jizba:2008:SuposPD,Hasegawa:2010:qExpBistable,Yalcin:2012:PolymerSS}.
The superstatistical stochastic process calculates a stationary distribution
$P_{st}(x)$ by assuming that the noise fluctuates over a long time
scale (i.e., very slowly): 
\begin{equation}
P_{st}(x)=\int_{0}^{\infty}dD\, P_{st}(x|D)P(D),\label{eq:superstat_def}
\end{equation}
where $P_{st}(x|D)$ is the stationary distribution given the noise
intensity $D$ and $P(D)$ is the distribution of $D$. Equation \ref{eq:superstat_def}
fits in the Bayesian framework by considering $P_{st}(x)$ and $P(D)$
as posterior and prior distributions, respectively.

We model temporal noise-intensity fluctuation dynamically and modulated
the intensity of white Gaussian noise by the Ornstein--Uhlenbeck process
in overdamped Langevin equations \cite{Hasegawa:2010:SIN,Hasegawa:2011:BistableSIN}:
\begin{eqnarray}
\dot{x} & = & -V^{\prime}(x)+s\xi_{x}(t),\label{eq:OD_Langevin_x}\\
\dot{s} & = & -\gamma(s-\alpha)+\sqrt{\gamma}\xi_{s}(t).\label{eq:OD_Langevin_s}
\end{eqnarray}
Here, $V(x)$ is the potential, $\alpha$ is the mean of the Ornstein--Uhlenbeck
process, $\gamma$ is the relaxation rate ($\gamma>0$), and $\xi_{x}(t)$
and $\xi_{s}(t)$ are white Gaussian noise with the correlation: 
\begin{eqnarray}
\left\langle \xi_{x}(t)\xi_{x}(t^{\prime})\right\rangle  & = & 2D_{x}\delta(t-t^{\prime}),\label{eq:Dx_def}\\
\left\langle \xi_{s}(t)\xi_{s}(t^{\prime})\right\rangle  & = & 2D_{s}\delta(t-t^{\prime}).\label{eq:Ds_def}
\end{eqnarray}
We call the term $s(t)\xi_{x}(t)$, the \emph{stochastic intensity
noise} (SIN) because the noise intensity is governed by a stochastic
process. SIN is the multiplicative term of white and colored noise,
and qualitatively different from white noise: it is in nonequilibrium.

In the context of Brownian transport, Reimann \textit{et al.}~\cite{Reimann:1996:OscBrownianMot}
first studied the transport effect with sinusoidal noise-intensity
modulation. Our work differs from this and succeeding studies that
employed a discrete dichotomous noise or a deterministic periodic
signal \cite{Reimann:1996:OscBrownianMot,Li:1997:Transport,Zhang:2008:OscTempRatchet};
in our model, fluctuations are governed by a continuous stochastic
process (the Ornstein--Uhlenbeck process). There exist similar models.
Borromeo \textit{et al.} studied a current generated by two symmetric
colored noises, the Ornstein--Uhlenbeck noise and its time-delayed
version, and observed the Maxwell's daemon-like phenomenon~\cite{Borromeo:2006:AutoMaxDeamon}.
Morgado \textit{et al.} investigated temporal heterogeneity in Poisson
mechanism~\cite{Morgado:2011:MassPoisson}. Our model focuses on
a multiplicative rather than an additive effect, because biological
phenomena are governed by multiplication (also see Discussion). This
distinction highlights the importance of SIN-induced transport.

In our calculations, we investigated the effect of four controllable
parameters on the current: 
\begin{itemize}
\item $\gamma$ (the relaxation rate in Equation~\ref{eq:UD_Langevin_s}), 
\item $Q$ (the effective noise intensity in Equation~\ref{eq:Q_def}), 
\item $\rho$ (the squared variation coefficient in Equation~\ref{eq:rho_def})
and 
\item $\mu$ (the scaled mass in Equation~\ref{eq:UD_Langevin_x}). 
\end{itemize}
Although $D_{x}$ (the noise-intensity in Equation~\ref{eq:UD_Langevin_x})
is also a controllable parameter, we kept it constant ($D_{x}=1$)
throughout the paper. The squared variation coefficient $\rho$, which
is generally defined as the ratio of the squared mean to the variance,
characterizes the deviation of the Gaussian noise distribution by
the kurtosis. The overdamped case is calculated using the matrix continued
fraction method (MCFM) and Monte Carlo (MC) simulations. The calculations
reveal that the current is a maximum at adequate $\gamma$ and $Q$.
This result concurs with resonant activation (RA) \cite{Doering:1992:ResonantActivation}
and noise-enhanced stability (NES) \cite{Mantegna:1996:NES,Spagnolo:2008:NES_review}
in the escape problem. Our main result is the enhanced transport capability
in intermediate time-correlation regions. This has an important biological
implication. The time-correlation of extrinsic noise (i.e., environmental
fluctuation) with gene expression in \emph{E. coli} is on the order
of cell cycle length~\cite{Rosenfeld:2005:SingleCell}. By fitting
the time-scale of our model to the \emph{E. coli} model in Ref.~\cite{Dunlop:2008:GeneNoiseCor},
we found that the time-correlation of extrinsic noise coincides with
the region of enhanced transport. It is a theoretical backup of the
tactful exploitation of environmental fluctuation by biological organisms.

\section{Methods}

\begin{figure}
\begin{centering}
\includegraphics[width=7cm]{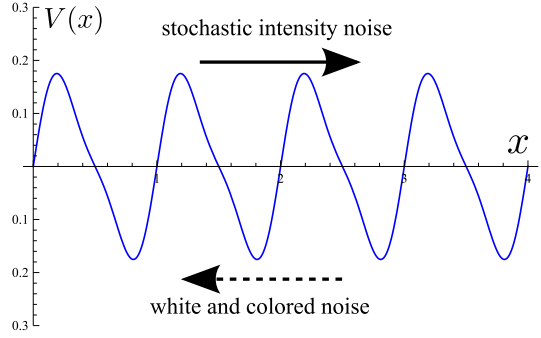} 
\par\end{centering}

\caption{(Online version in color) A ratchet potential of Equation~\ref{eq:potential}.
The dashed and solid arrows indicate the directions of the current
in the correlation ratchet (driven by additive white and colored noise)
and the ratchet (driven by SIN), respectively. \label{fig:potential}}
\end{figure}

\begin{figure}
\begin{centering}
\includegraphics[width=12cm]{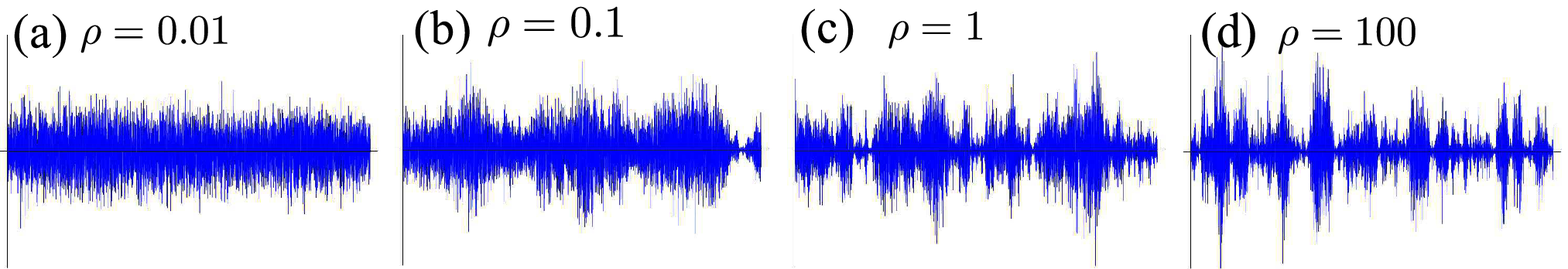} 
\par\end{centering}

\begin{centering}
\includegraphics[width=12cm]{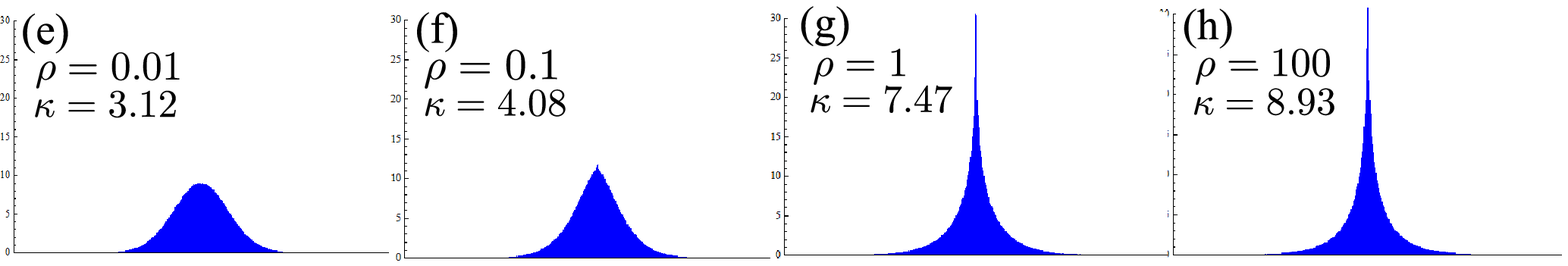} 
\par\end{centering}

\caption{(Online version in color) (a)--(d) Time courses of SIN generated by
MC simulations for four values of $\rho$ (squared variation coefficient
given by Equation~\ref{eq:rho_def}): $\rho=$ (a) 0.01, (b) 0.1,
(c) 1, and (d) 100. (e)--(h) Histograms of SIN for the four values
of $\rho$ and their sample kurtosis $\kappa$. In (a)--(h), we varied
$\rho$ while keeping the effective intensity $Q=D_{x}(D_{s}+\alpha^{2})$
constant. \label{fig:example_traj}}
\end{figure}

\subsection{Brownian Particles}

Brownian particles are subject to noise-intensity fluctuations represented
by 
\begin{eqnarray}
\mu\ddot{x} & = & -\dot{x}-V^{\prime}(x)+s\xi_{x}(t),\label{eq:UD_Langevin_x}\\
\dot{s} & = & -\gamma(s-\alpha)+\sqrt{\gamma}\xi_{s}(t),\label{eq:UD_Langevin_s}
\end{eqnarray}
where the scaled mass $\mu$ is introduced into Equation~\ref{eq:OD_Langevin_x}
for a study of a mass separation effect, and meanings of $V(x)$ are
the same as in Equation~\ref{eq:OD_Langevin_x}. We used the same
ratchet potential function as used in previous studies \cite{Bartussek:1996:CorRatchet,Lindner:1999:InertiaRatchet}
\begin{equation}
V(x)=\frac{1}{2\pi}\left\{ \sin(2\pi x)+\frac{1}{4}\sin(4\pi x)\right\} +Fx,\label{eq:potential}
\end{equation}
with a periodicity $V(x+1)=V(x)$ ($F=0$), where $F$ is the load.
Figure~\ref{fig:potential} shows the potential with no load ($F=0$),
where the dashed and solid arrows indicate the normal current direction
for the correlation ratchet \cite{Bartussek:1996:CorRatchet,Lindner:1999:InertiaRatchet}
and for the ratchet driven by SIN, respectively. The current direction
for the SIN case is identical to that in Ref.~\cite{Li:1997:Transport}
in which the noise intensity is modulated by a random dichotomous
process. In Equation~\ref{eq:UD_Langevin_s}, the relaxation rate
$\gamma$ denotes the inverse of a time scale (time-correlation).
When the noise intensity fluctuates with a longer time scale ($\gamma\rightarrow0$),
systems driven by SIN locally equilibrates and hence a current is
not generated in accordance with the second law of thermodynamics.
Likewise, SIN reduces to white noise (with a noise intensity $Q$)
when the noise-intensity fluctuates very rapidly ($\gamma\rightarrow\infty$)
\cite{Hasegawa:2010:SIN}, which also indicates that the current vanishes.

\begin{figure}
\begin{centering}
\includegraphics[width=5.5cm]{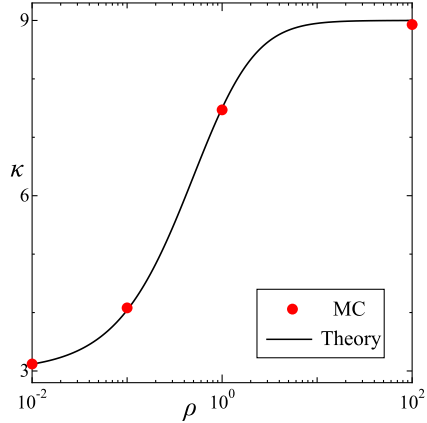} 
\par\end{centering}

\caption{(Online version in color) Kurtosis $\kappa$ of SIN as a function
of $\rho$. The solid line indicates Equation~\ref{eq:kurtosis_rho}
and the filled circles indicate the sample kurtosis in Figures~\ref{fig:example_traj}(e)--(h).
\label{fig:kurtosis_plo}}
\end{figure}

For Equation~\ref{eq:UD_Langevin_s}, the stationary distribution
$P_{st}(s)$ of the intensity-modulation term $s$ is given by 
\begin{equation}
P_{st}(s)=\frac{1}{\sqrt{2\pi D_{s}}}\exp\left\{ -\frac{1}{2D_{s}}(s-\alpha)^{2}\right\} .\label{eq:statdist_s}
\end{equation}
A calculation of the correlation function of SIN yields \cite{Hasegawa:2011:BistableSIN}
\begin{equation}
\left\langle s(t)\xi_{x}(t)s(t^{\prime})\xi_{x}(t^{\prime})\right\rangle =2Q\delta(t-t^{\prime}),\label{eq:SIN_correlation_func}
\end{equation}
with 
\begin{equation}
Q=D_{x}(D_{s}+\alpha^{2}),\label{eq:Q_def}
\end{equation}
where $Q$ expresses the effective noise intensity and $D_{x}$ and
$D_{s}$ are the noise intensities of $\xi_{x}(t)$ and $\xi_{s}(t)$,
respectively {[}Equations~\ref{eq:Dx_def} and \ref{eq:Ds_def}{]}.
Here, we introduce the squared variation coefficient $\rho$ of the
noise-intensity fluctuations \cite{Hasegawa:2011:BistableSIN}: 
\begin{equation}
\rho=D_{s}/\alpha^{2},\label{eq:rho_def}
\end{equation}
which denotes the squared ratio of the standard deviation to the mean
of Equation~\ref{eq:statdist_s}. Figures~\ref{fig:example_traj}(a)--(d)
show trajectories of SIN with $\rho=$ 0.01, 0.1, 1, and 100, respectively.
All share the same effective noise intensity $Q$ (the same variance).
For $\rho=0$, SIN reduces to white Gaussian noise with a noise intensity
$Q=D_{x}\alpha^{2}$. On increasing $\rho$ (Figures~\ref{fig:example_traj}(e)--(h)),
the noise-intensity fluctuations become larger and the distribution
of SIN deviates from the Gaussian distribution. To quantify the deviation,
we calculate the kurtosis of SIN, which is a measure of heavy tails
in probability density functions. The kurtosis $\kappa$ of SIN (i.e.,
$s(t)\xi_{x}(t)$) is given by (Appendix~\ref{sec:kurtosis}) 
\begin{equation}
\kappa=\frac{\left\langle \left\{ s(t)\xi_{x}(t)\right\} ^{4}\right\rangle }{\left\langle \left\{ s(t)\xi_{x}(t)\right\} ^{2}\right\rangle ^{2}}=9-\frac{6}{(1+\rho)^{2}}\hspace{1em}(0\le\rho<\infty),\label{eq:kurtosis_rho}
\end{equation}
where $\kappa$ depends only on the squared variation coefficient
$\rho$ having a crucial effect on the statistical properties of SIN.
Equation~\ref{eq:kurtosis_rho} shows that the kurtosis is 3 for
$\rho=0$ and $\kappa$ increases with increasing $\rho$, giving
a flatter distribution. Equation~\ref{eq:kurtosis_rho} is plotted
by the solid curve in Figure~\ref{fig:kurtosis_plo} in which the
filled circles denote the kurtosis calculated by MC simulations.

\section{Results}

In Equations~\ref{eq:UD_Langevin_x} and \ref{eq:UD_Langevin_s},
our model has four parameters ($\gamma$, $\alpha$, $D_{x}$, and
$D_{s}$) in terms of noise properties. Instead of these parameters,
we adopt $\gamma$, $\rho$ (the squared variation coefficient in
Equation~\ref{eq:rho_def}), $Q$ (the effective noise intensity
in Equation~\ref{eq:Q_def}), and $D_{x}$ as controllable parameters
in model calculations, where $Q$ and $\rho$ are defined by Equations~\ref{eq:Q_def}
and \ref{eq:rho_def}, respectively. The new parameters ($\gamma$,
$\rho$, $Q$, and $D_{x}$) specify the system identically since
$\alpha$ and $D_{s}$ are uniquely determined by $D_{x}$, $Q$,
and $\rho$: 
\begin{equation}
\alpha=\sqrt{\frac{Q}{D_{x}(1+\rho)}},\hspace{1em}D_{s}=\frac{\rho Q}{D_{x}(1+\rho)}.\label{eq:alpha_Ds_def}
\end{equation}
Performing model calculations based on the MCFM and MC simulations,
we studied the dependence of the current on $\gamma$, $\rho$, and
$Q$ in the overdamped case ($D_{x}$ is set to $D_{x}=1$ throughout
the paper). In the underdamped case, we additionally investigated
the dependence of the current on the scaled mass $\mu$ (Equation~\ref{eq:UD_Langevin_x})
with MC simulations. MC simulations were performed with the Euler-forward
method with a time resolution of $\Delta t=10^{-4}$ (for details
of the method, see Ref.~\cite{Risken:1989:FPEBook}).

\subsection{Overdamped case}

We first calculated the current $J$ for the overdamped case ($\mu=0$
in Equation~\ref{eq:UD_Langevin_x}). For $t\rightarrow\infty$,
the stationary distribution $P_{st}(x,s)$ of $(x,s)$ has to satisfy
the stationary Fokker--Planck equation (FPE) 
\begin{equation}
\mathbb{L}_{\mathrm{FP}}P_{st}(x,s)=0,\label{eq:FPE_OD}
\end{equation}
where $\mathbb{L}_{\mathrm{FP}}$ is an FPE operator: 
\begin{equation}
\mathbb{L}_{\mathrm{FP}}=\frac{\partial}{\partial x}V^{\prime}(x)+D_{x}\frac{\partial^{2}}{\partial x^{2}}s^{2}+\gamma\left\{ \frac{\partial}{\partial s}(s-\alpha)+D_{s}\frac{\partial^{2}}{\partial s^{2}}\right\} .\label{eq:FPE_op_over}
\end{equation}
Since Equation~\ref{eq:FPE_OD} is not written in terms of potential
forms, we cannot calculate the stationary distribution in a closed
form. Consequently, we used the MCFM to solve Equation~\ref{eq:FPE_OD},
which expands $P_{st}(x,s)$ in terms of a complete orthonormal set.
The MCFM is a common technique for stochastic processes and it is
widely used to solve FPEs (see Ref.~\cite{Risken:1989:FPEBook} and
references therein). Considering the periodicity of the potential
{[}$V^{\prime}(x+1)=V^{\prime}(x)${]} and domain {[}$x\in(-\infty,\infty)$
and $s\in(-\infty,\infty)${]}, we expanded the stationary distribution
$P_{st}(x,s)$ in a Fourier series for $x$ and the Hermite function
for $s$: 
\begin{equation}
P_{st}(x,s)=\psi_{0}(s)\sum_{k=-M_{k}}^{M_{k}}\sum_{n=0}^{M_{n}}C_{k,n}\exp\left(2\pi k\mathrm{i}x\right)\psi_{n}(s).\label{eq:statdist_overdamped}
\end{equation}
Here, $C_{k,n}$ are expansion coefficients, $M_{k}$ and $M_{n}$
are truncation numbers on which the precision of obtained solutions
depends, and $\psi_{n}(s)$ is the Hermite function satisfying the
orthonormality relation $\left\langle \psi_{n^{\prime}}(s)\psi_{n}(s)\right\rangle =\delta_{n^{\prime}n}$:
\begin{eqnarray}
\psi_{n}(s) & = & \left(\frac{1}{2\pi D_{s}}\right)^{1/4}\sqrt{\frac{1}{2^{n}n!}}H_{n}(\eta)\exp\left(-\frac{1}{2}\eta^{2}\right)\hspace{1em}\eta=\sqrt{\frac{1}{2D_{s}}}(s-\alpha),\label{eq:Hermite_s}
\end{eqnarray}
where $H_{n}(s)$ is the $n$th Hermite polynomial. Multiplying Equation~\ref{eq:FPE_OD}
by $\exp(-2\pi k^{\prime}\mathrm{i}x)\psi_{n^{\prime}}(s)/\psi_{0}(s)$
and integrating over $x$ and $s$, we obtain a linear algebraic equation
in terms of $C_{k,n}$, which can be solved by the MCFM (Appendix~\ref{sec:MCFM}).
The current $J$ is calculated using 
\begin{equation}
J=\left\langle v\right\rangle =\int_{0}^{1}dx\left(\int_{-\infty}^{\infty}ds\, J_{x}(x,s)\right),\label{eq:over_current}
\end{equation}
where $J_{x}(x,s)$ is the probability current in the $x$ direction
due to the continuity equation $\partial_{t}P+\partial_{x}J_{x}+\partial_{s}J_{s}=0$:
\begin{equation}
J_{x}(x,s)=\left(-V^{\prime}(x)-D_{x}s^{2}\frac{\partial}{\partial x}\right)P_{st}(x,s).\label{eq:J_x}
\end{equation}
By substituting Equation~\ref{eq:statdist_overdamped} into Equation~\ref{eq:over_current},
the current $J$ can be expressed in terms of $C_{k,n}$: 
\begin{equation}
J=-\frac{1}{2}(C_{-1,0}+C_{1,0})-\frac{1}{4}\left(C_{-2,0}+C_{2,0}\right)-FC_{0,0}.\label{eq:over_current_C}
\end{equation}
The current given by Equation~\ref{eq:over_current_C} does not vanish
even when there is no load (i.e. $F=0$) due to the broken detailed
balance.

\begin{figure}
\begin{centering}
\includegraphics[width=7cm]{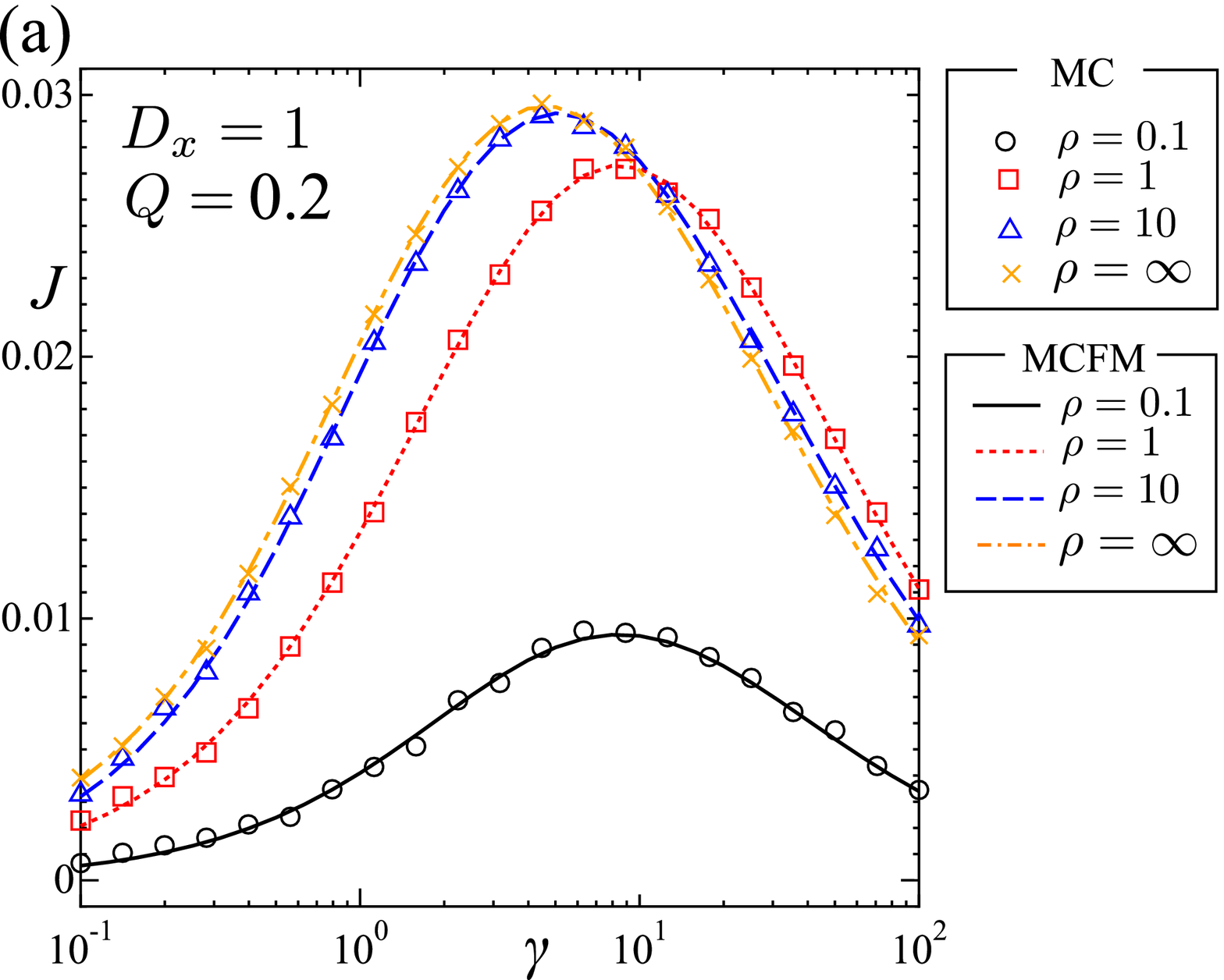}~~~~\includegraphics[width=7cm]{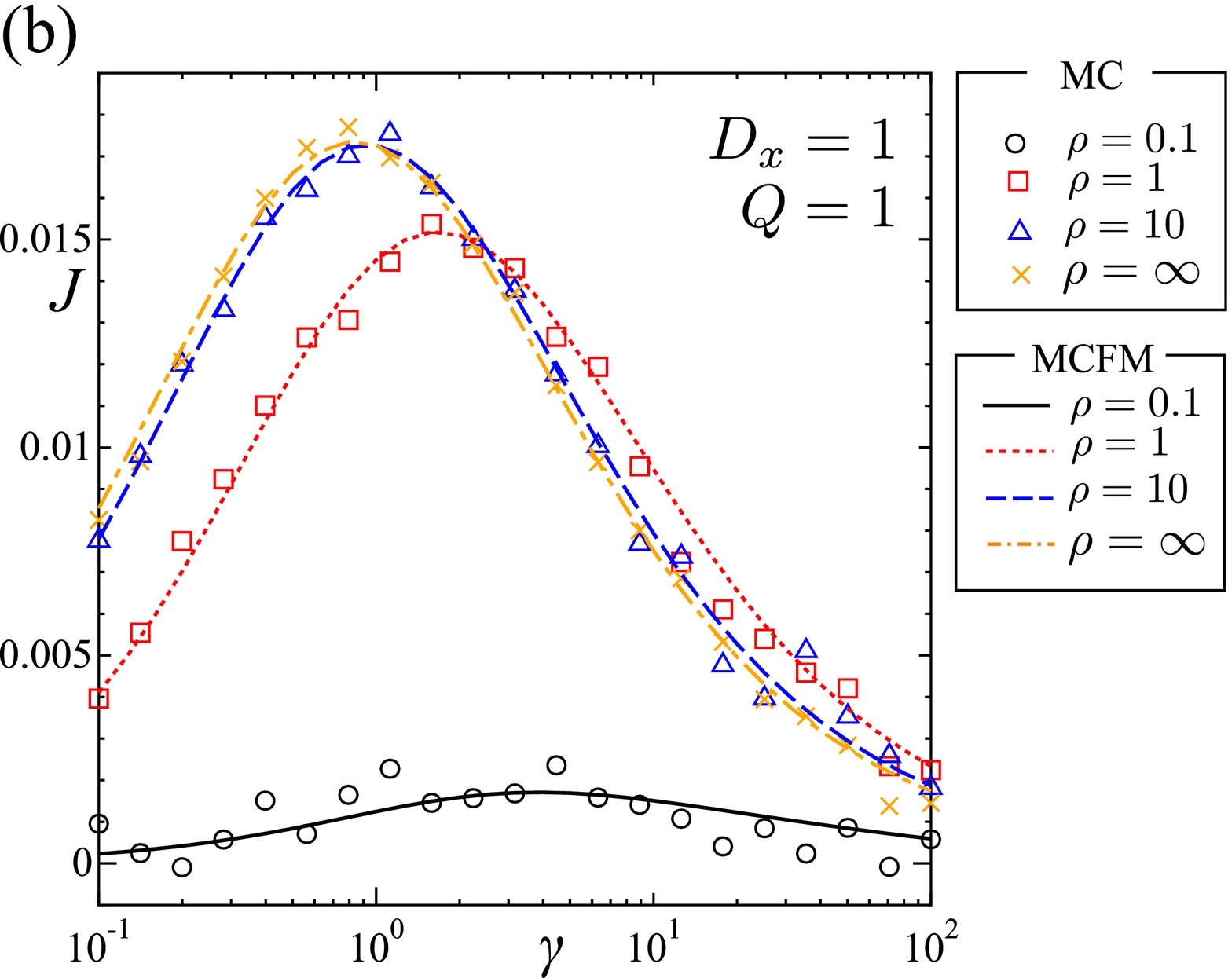} 
\par\end{centering}

\caption{(Online version in color) Current $J$ as a function $\gamma$ for
the overdamped case ($\mu=0$). The lines and symbols represent $J$
calculated using the MCFM and MC simulations, respectively. The parameters
are (a) $D_{x}=1$ and $Q=0.2$ and (b) $D_{x}=1$ and $Q=1$, with
$\rho=0.1$ (solid lines and circles), $\rho=1$ (dotted lines and
squares), $\rho=10$ (dashed lines and triangles), and $\rho=\infty$
(dot-dashed lines and crosses). \label{fig:OD_v_gamma}}
\end{figure}

\begin{figure}
\begin{centering}
\includegraphics[width=7cm]{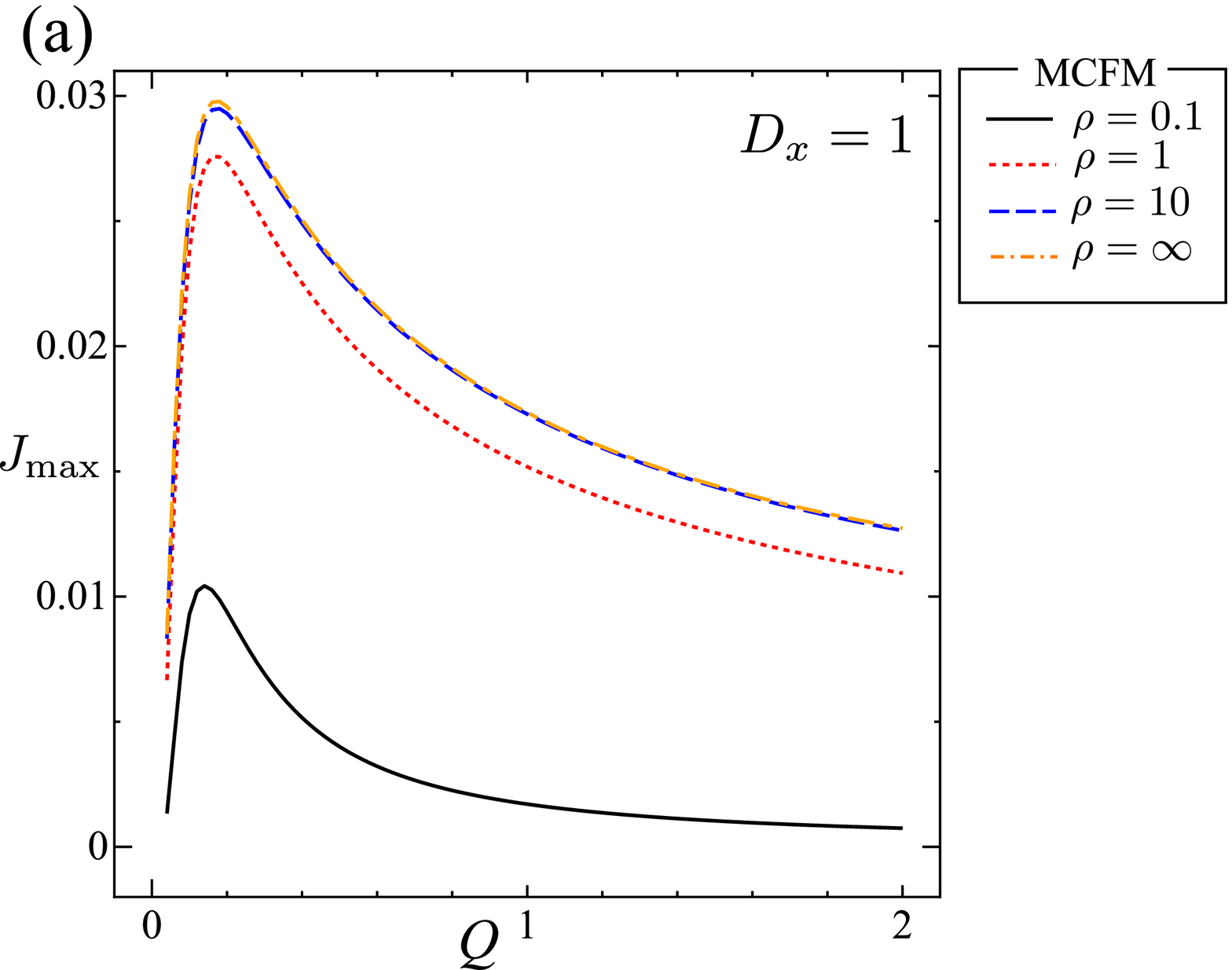}~~~~\includegraphics[width=7cm]{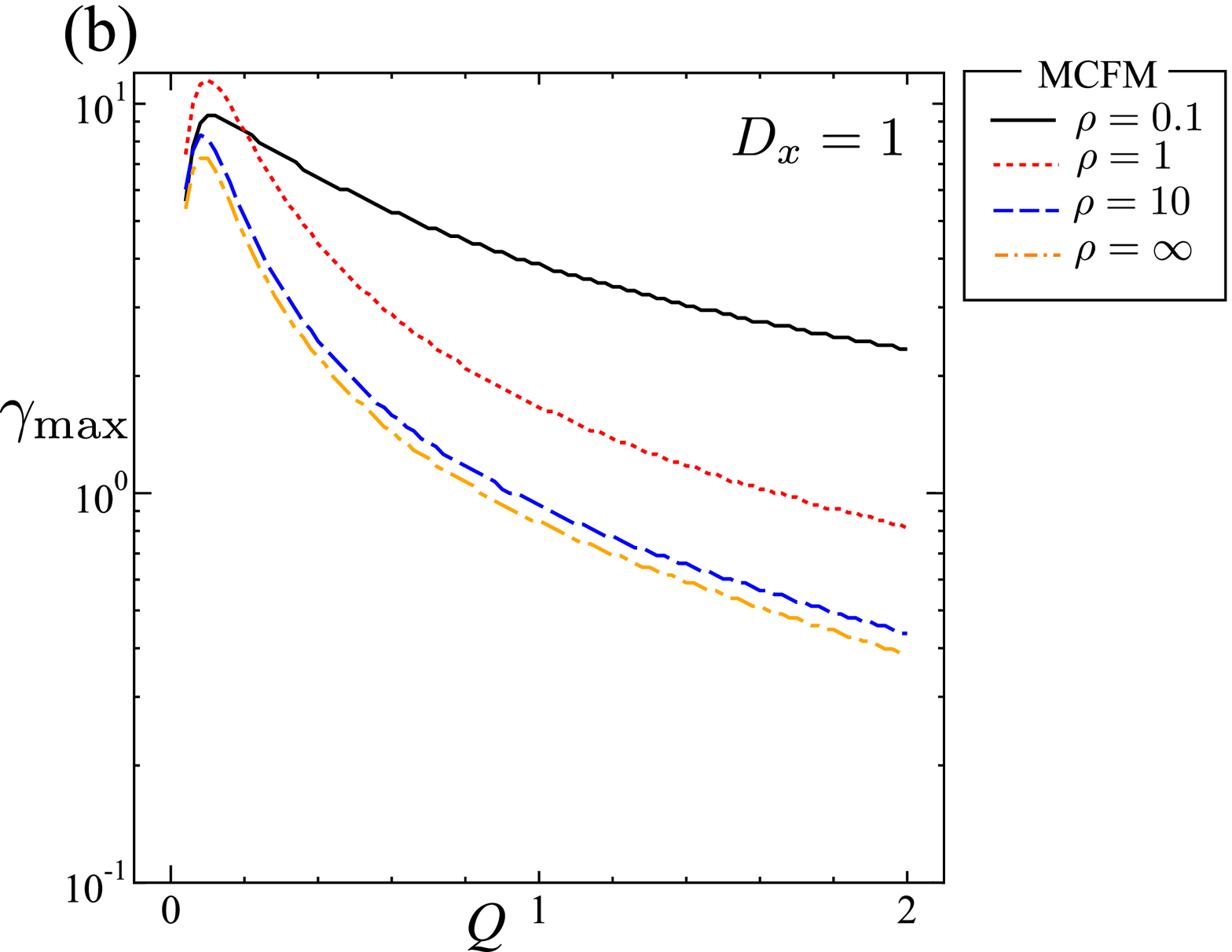} 
\par\end{centering}

\caption{(Online version in color) (a) Current $J_{\max}$ and (b) $\gamma_{\max}$
as a function $Q$ for the overdamped case ($\mu=0$). The parameters
are $D_{x}=1$ with $\rho=0.1$ (solid lines), $\rho=1$ (dotted lines),
$\rho=10$ (dashed lines), and $\rho=\infty$ (dot-dashed lines).
\label{fig:OD_v_Q}}
\end{figure}

\begin{figure}
\begin{centering}
\includegraphics[width=7cm]{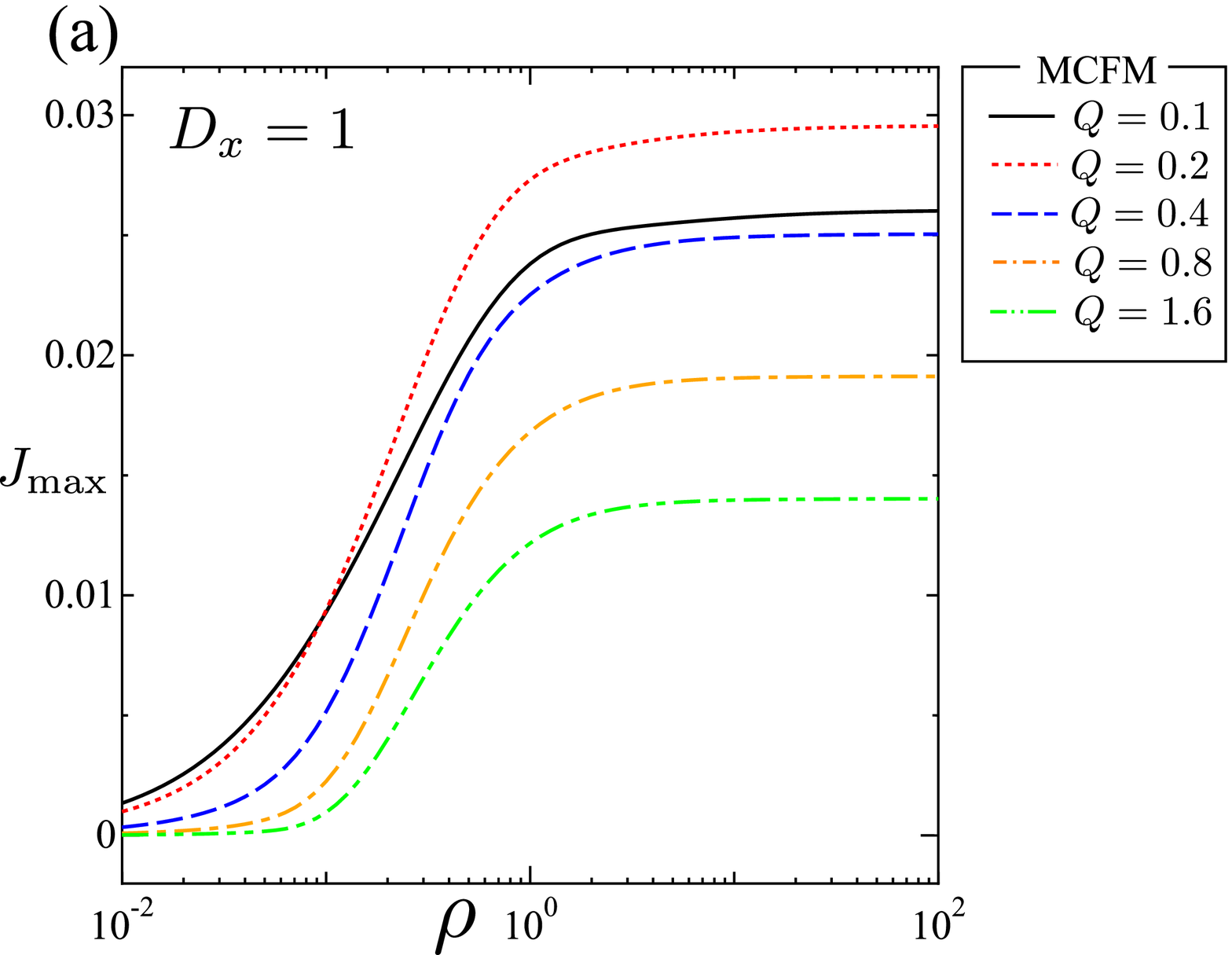}~~~~\includegraphics[width=7cm]{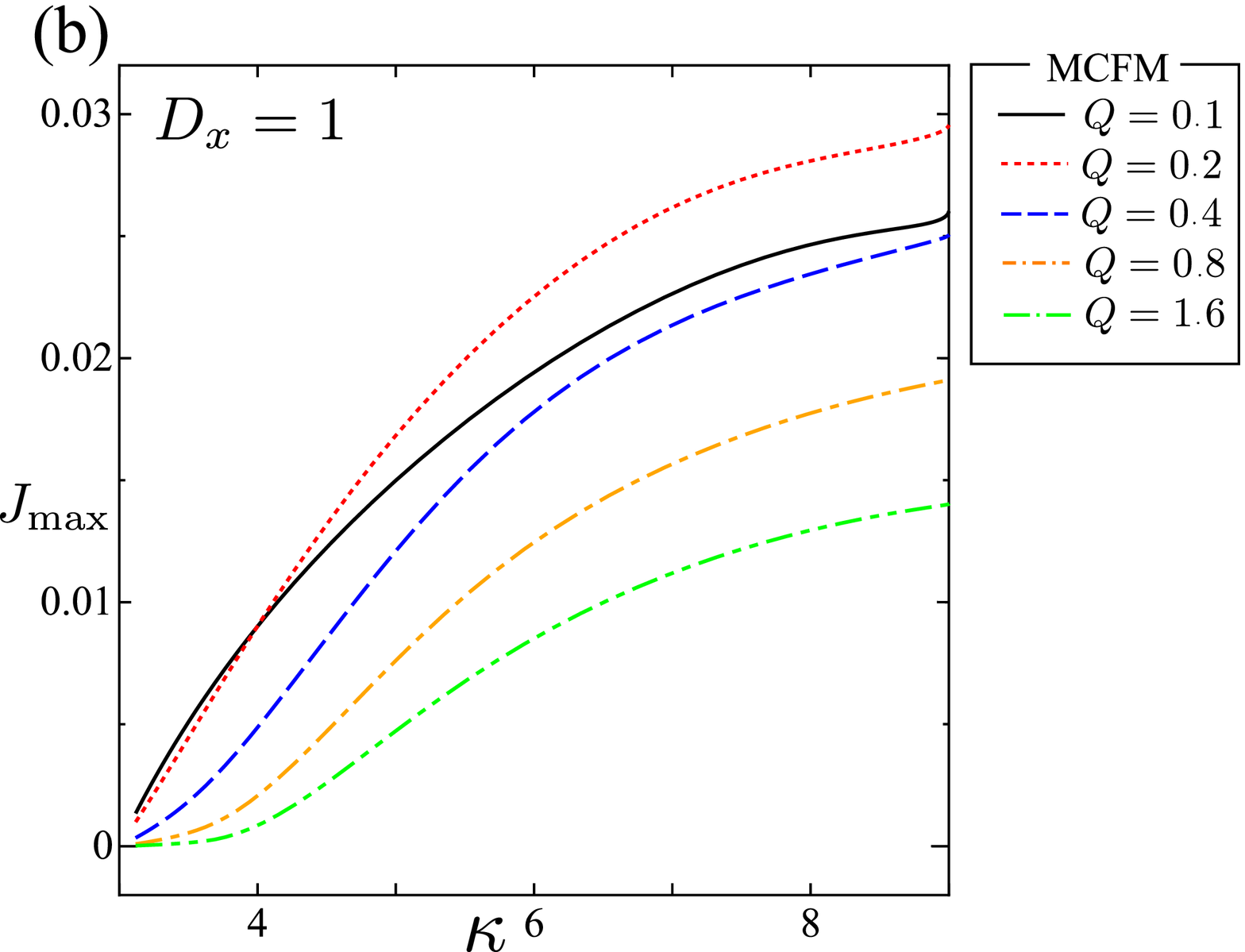} 
\par\end{centering}

\caption{(Online version in color) $J_{\max}$ as a function of (a) squared
variation coefficient $\rho$ and (b) kurtosis $\kappa$ for the overdamped
case ($\mu=0$). The parameters are $D_{x}=1$ with $Q=0.1$ (solid
lines), $Q=0.2$ (dotted lines), $Q=0.4$ (dashed lines), $Q=0.8$
(dot-dashed lines), and $Q=1.6$ (dot-dot-dashed lines).\label{fig:OD_v_rho}}
\end{figure}

In practical calculations of Equation~\ref{eq:over_current_C}, we
increased the truncation numbers $M_{k}$ and $M_{n}$ in Equation~\ref{eq:statdist_overdamped}
until the current $J$ converged. We also performed MC simulations
to verify reliability of the MCFM, where the velocity is given by
$v=[x(T)-x(0)]/T$ ($T=10^{5}$). The calculation was repeated $100$
times and the average velocity was determined. Below, we calculate
the dependence of the current $J$ on the relaxation rate $\gamma$
(Figure~\ref{fig:OD_v_gamma}), the effective noise intensity $Q$
(Figure~\ref{fig:OD_v_Q}), and the squared variation coefficient
$\rho$ (Figure~\ref{fig:OD_v_rho}). 

We first show the current $J$ as a function of the relaxation rate
$\gamma$ {[}Equation~\ref{eq:OD_Langevin_s}{]}. In Figure~\ref{fig:OD_v_gamma},
$\gamma$ is plotted against the current $J$ for two sets of parameters:
Case (a) $D_{x}=1$ and $Q=0.2$ (Figure~\ref{fig:OD_v_gamma}(a)),
and Case (b) $D_{x}=1$ and $Q=1$ (Figure~\ref{fig:OD_v_gamma}(b)).
The relationship was computed for four values of $\rho$ (the squared
variation coefficient): $\rho=0.1,1,10$, and $\infty$ (the case
$\rho=\infty$ corresponds to $\alpha=0$ and $D_{s}=Q/D_{x}$). In
Case (a), $J$ became a maximum at an intermediate $\gamma$, while
the current vanished for both $\gamma\rightarrow0$ and $\gamma\rightarrow\infty$.
The small current at $\rho=0.1$ was expected because a small $\rho$
corresponds to weak noise fluctuations. Because larger $\rho$ means
larger fluctuations (the kurtosis monotonically increases as a function
of $\rho$, as shown in Figure~\ref{fig:kurtosis_plo}), it is natural
that the current increases as $\rho$ increases. However, $J$ in
the range $\gamma>10$ shows a different tendency and it is higher
when $\rho=1$ than when $\rho=10$ or $\rho=\infty$. This is due
to the noise-intensity modeling in Equation~\ref{eq:UD_Langevin_s};
specifically, $s(t)$ in Equation~\ref{eq:UD_Langevin_s} for small
$\rho$ rarely has negative values, whereas $s(t)$ for large $\rho$
has negative as well as positive values. For large $\gamma$ cases,
the ``effective'' relaxation rate might be measured by $\left\langle |s(t)||s(t^{\prime})|\right\rangle $
rather than by $\left\langle s(t)s(t^{\prime})\right\rangle $, which
indicates that the effective relaxation rates for larger $\rho$ cases
are larger than their actual values. In Figure~\ref{fig:OD_v_gamma},
the results obtained by the MCFM (lines) are always in agreement with
MC simulations (symbols), which indicates the reliability of the MCFM.
The MC simulations tended to converge for smaller $Q$ cases because
$v$ is more centered on the mean (thin distributions) when the effective
noise intensity is smaller. We also note the difference between our
model and that of Ref.~\cite{Reimann:1996:OscBrownianMot} where
noise-intensity is modulated by a deterministic sinusoidal signal.
We found that the width of peaks tends to be wider in our model, which
is a consequence that the power spectrum of Ornstein--Uhlenbeck noise
have the Lorentzian function whereas that of a sinusoidal function
is a delta-peaked function.

We next show the current $J$ as a function of the effective noise
intensity $Q$ {[}Equation~\ref{eq:Q_def}{]}. Since the effect of
the relaxation rate $\gamma$ varies depending on the squared variation
coefficient $\rho$ (as shown above), we removed the $\gamma$ dependence
of the current $J$ by taking the maximum in terms of $\gamma$:
\begin{eqnarray}
J_{\max}=J_{\max}(D_{x},Q,\rho) & = & \underset{\gamma}{\max}\, J(D_{x},\gamma,Q,\rho),\label{eq:J_max_def}\\
\gamma_{\max}=\gamma_{\max}(D_{x},Q,\rho) & = & \underset{\gamma}{\mathrm{argmax}}\, J(D_{x},\gamma,Q,\rho).\label{eq:gamma_max_def}
\end{eqnarray}
Here, $J_{\max}$ and $\gamma_{\max}$ respectively denote the maximum
current and $\gamma$ when the current is a maximum. Figures~\ref{fig:OD_v_Q}(a)
and (b) show plots of $Q$ against $J_{\max}$ and $\gamma_{\max}$
for $D_{x}=1$, respectively. Again, computations were performed for
four values of $\rho$. In Figure~\ref{fig:OD_v_Q}(a), $J_{\max}$
has maxima as a function $Q$; the magnitude of this maxima is small
for $\rho=0.1$, as in Figure~\ref{fig:OD_v_gamma}. Transport effects
are robust in terms of the noise intensity $Q$, because the current
for $Q=2$ still exhibits half the maximum current (around $Q=0.2$)
for large $\rho$ cases. Figure~\ref{fig:OD_v_Q}(b) shows $\gamma_{\max}$
(i.e., the relaxation rate yielding the maximum current) as a function
of $Q$. For all four values of $\rho$, $\gamma_{\max}$ has a maximum
value around $Q=0.1$ and decreases with increasing $Q$. This result
shows that the noise intensity has to fluctuate over a long time scale
to enhance the transport capability in noisy environments.

We also investigated the dependence of $\rho$ on $J_{\max}$ while
keeping the other parameters constant. Figure~\ref{fig:OD_v_rho}(a)
shows $J_{\max}$ (Equation~\ref{eq:J_max_def}) as a function of
$\rho$ with $D_{x}=1$ and five values of $Q$ (the effective noise
intensity): $Q=0.1,0.2,0.4,0.8$, and $1.6$. $J_{\max}$ increases
monotonically; the increase exhibits sigmoid-like behavior in terms
of $\log\rho$, indicating that $J_{\max}$ is an extremely nonlinear
function in terms of $\rho$. Figure~\ref{fig:OD_v_rho}(b) shows
the dependence of $J_{\max}$ on the kurtosis $\kappa$ (Equation~\ref{eq:kurtosis_rho}).
Although $J_{\max}$ still exhibits nonlinearity as a function of
$\kappa$, its nonlinearity is much smaller than that of $\rho$.
This result shows that the kurtosis can be used as an index parameter
for Brownian transport driven by noise-intensity fluctuations.

\subsection{Underdamped case}

We next investigated Brownian transport (Equations~\ref{eq:UD_Langevin_x}
and \ref{eq:UD_Langevin_s}) in the underdamped regime, especially
from the viewpoint of a mass separation effect \cite{Marchesoni:1998:MassSep}.
Although we could use the MCFM for the overdamped case, the MCFM for
an underdamped FPE did not yield stable solutions in terms of the
current (data not shown) so that we used MC simulations only. The
underdamped model includes an extra parameter $\mu$ (scaled mass)
in addition to the four parameters ($\gamma$, $\rho$, $Q$, and
$D_{x}$). The velocity is given by $v=[x(T)-x(0)]/T$ ($T=10^{5}$).
The calculation was repeated $100$ times to obtain the current as
the average velocity.

\begin{figure}
\begin{centering}
\includegraphics[width=7cm]{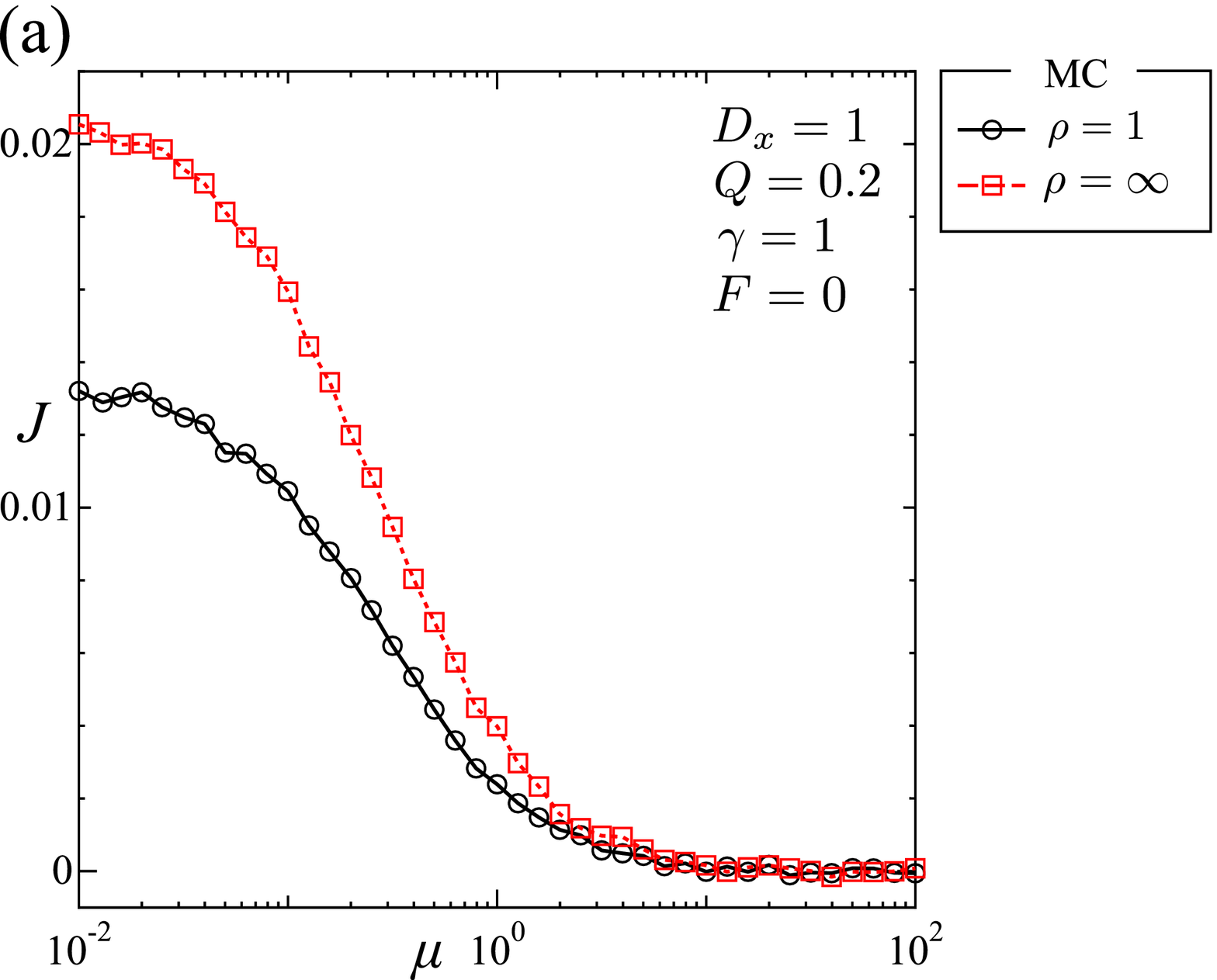}~~~~\includegraphics[width=7cm]{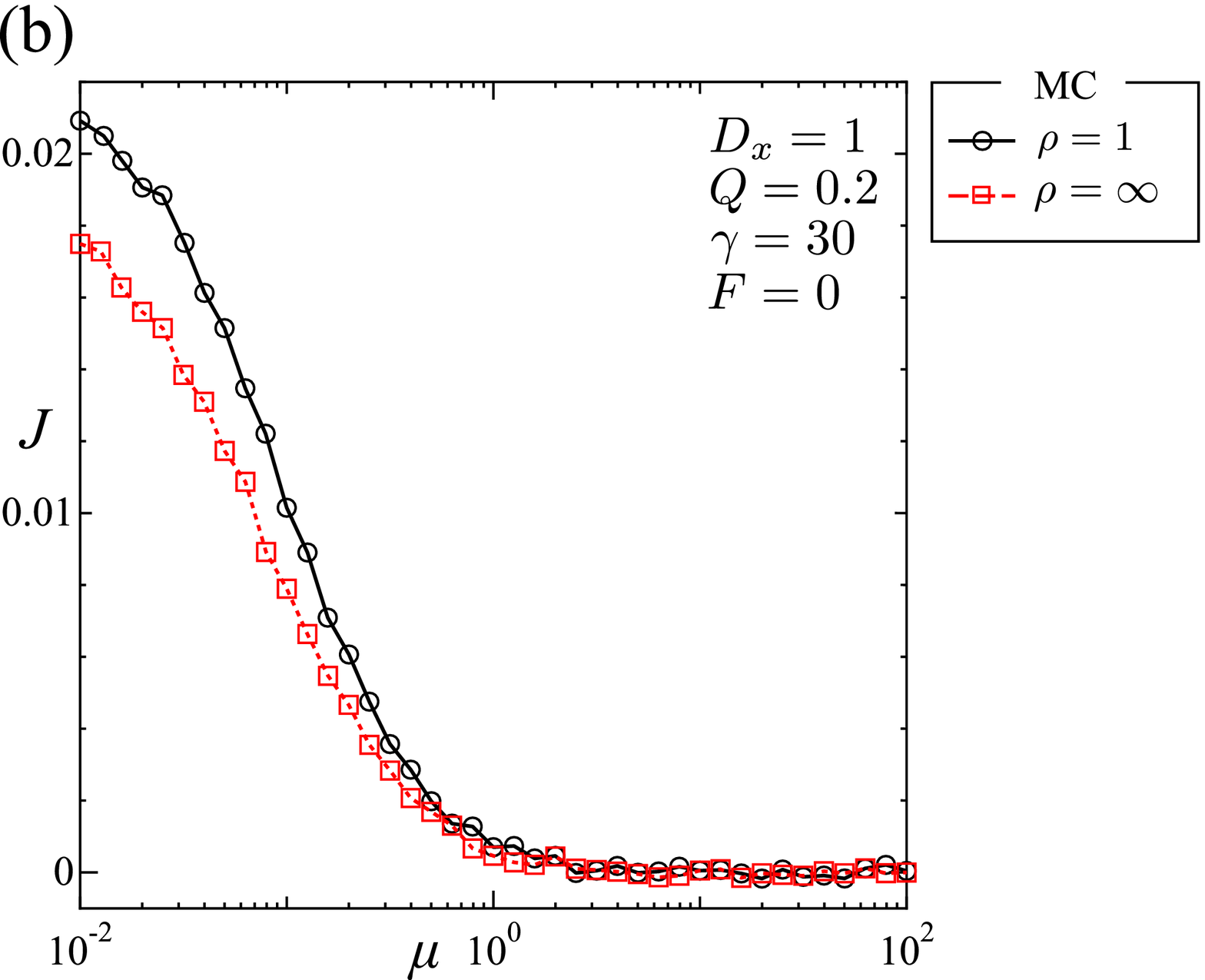} 
\par\end{centering}

\caption{(Online version in color) Current $J$ as a function of $\mu$ using
MC simulations for (a) $\gamma=1$ and (b) $\gamma=30$. The other
parameters are $D_{x}=1$, $Q=0.2$, and $F=0$ with $\rho=1$ (circles)
and $\rho=\infty$ (squares). Lines are included as a guide to the
eye only.\label{fig:J_as_mass}}
\end{figure}

We first show the current $J$ as a function of the scaled mass $\mu$.
Figure~\ref{fig:J_as_mass} shows the dependence of the current $J$
on $\mu$ for two values of $\gamma$ {[}Case (a) $\gamma=1$ (Figure~\ref{fig:J_as_mass}(a))
and Case (b) $\gamma=30$ (Figure~\ref{fig:J_as_mass}(b)){]}, where
the other parameters are $D_{x}=1$, $Q=0.2$, $F=0$, and $\rho=1$
(circles) or $\rho=\infty$ (squares). In Case (a), the current decreases
monotonically and the current for $\rho=\infty$ always exceeds that
for $\rho=1$. In contrast, in Case (b), the current of $\rho=\infty$
is always smaller than that of $\rho=1$ and the relation between
the magnitudes for $\rho=1$ and $\rho=\infty$ differs from that
for Case (a). As shown in Figure~\ref{fig:OD_v_Q}, the ``effective''
relaxation rate for larger $\rho$ is larger than the actual relaxation
rate. This is the reason why converse magnitude relaxation occurs
between $\rho=1$ and $\rho=\infty$ in Case (b).

\begin{figure}
\begin{centering}
\includegraphics[width=7cm]{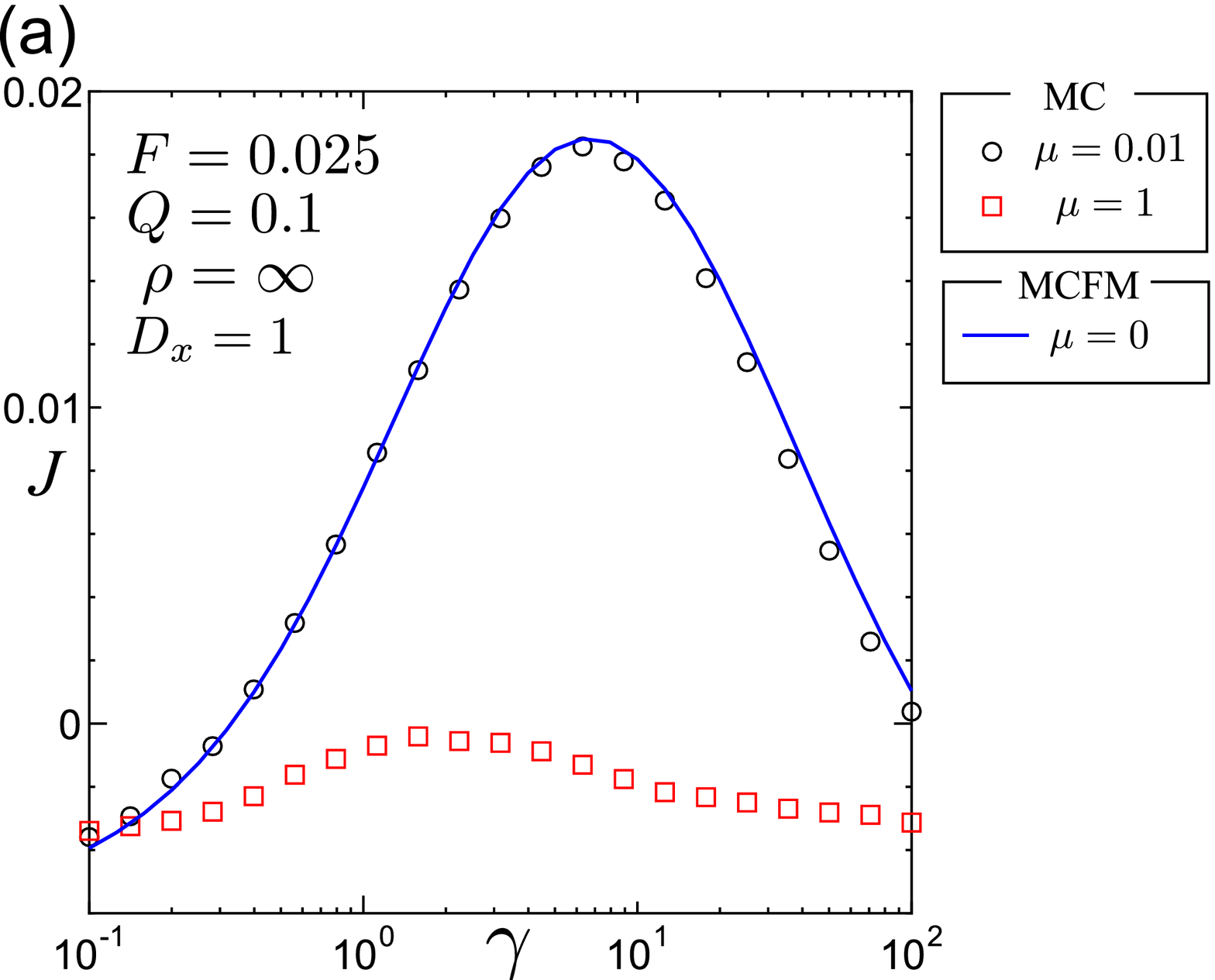}~~~~\includegraphics[width=7cm]{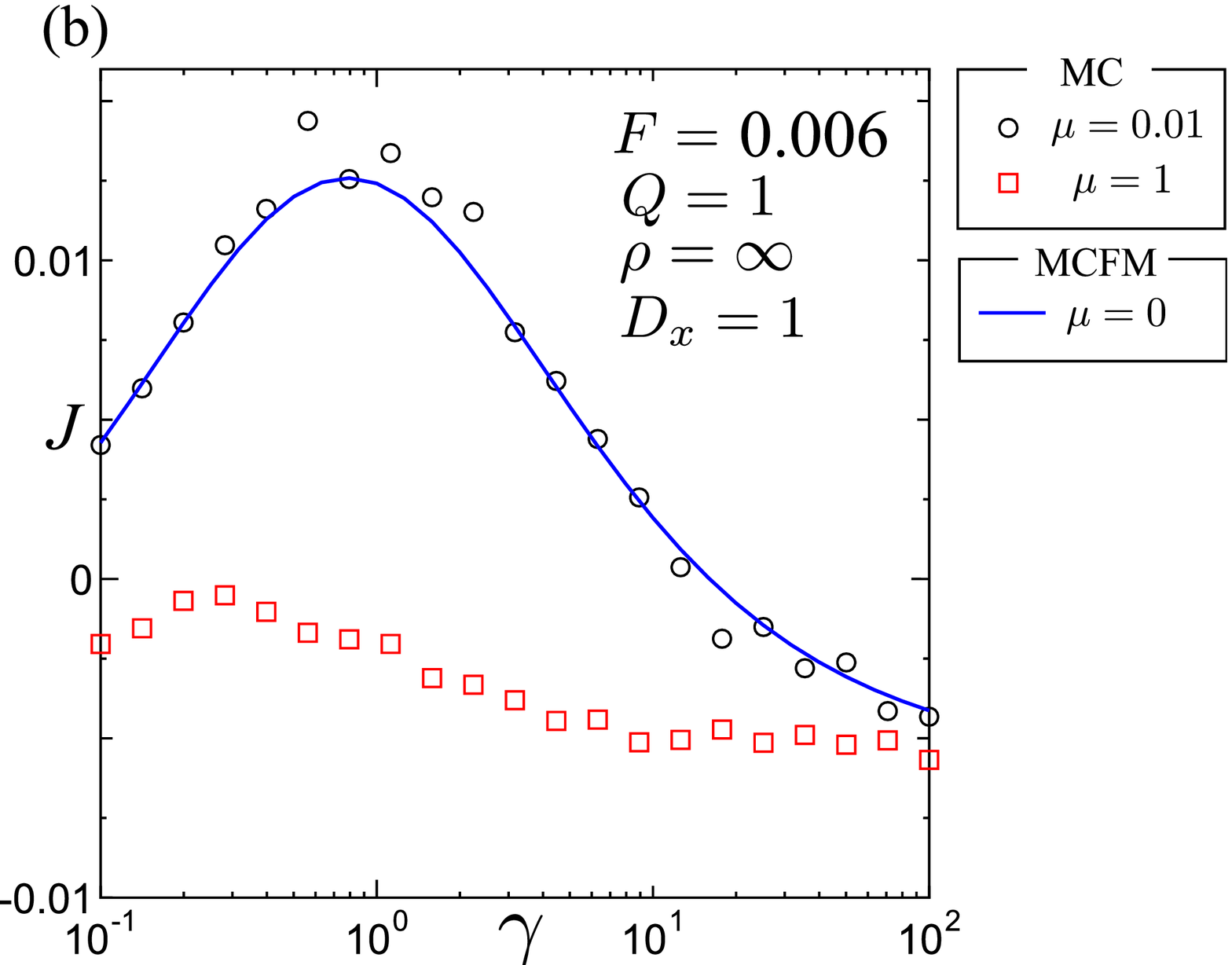} 
\par\end{centering}

\begin{centering}
\includegraphics[width=7cm]{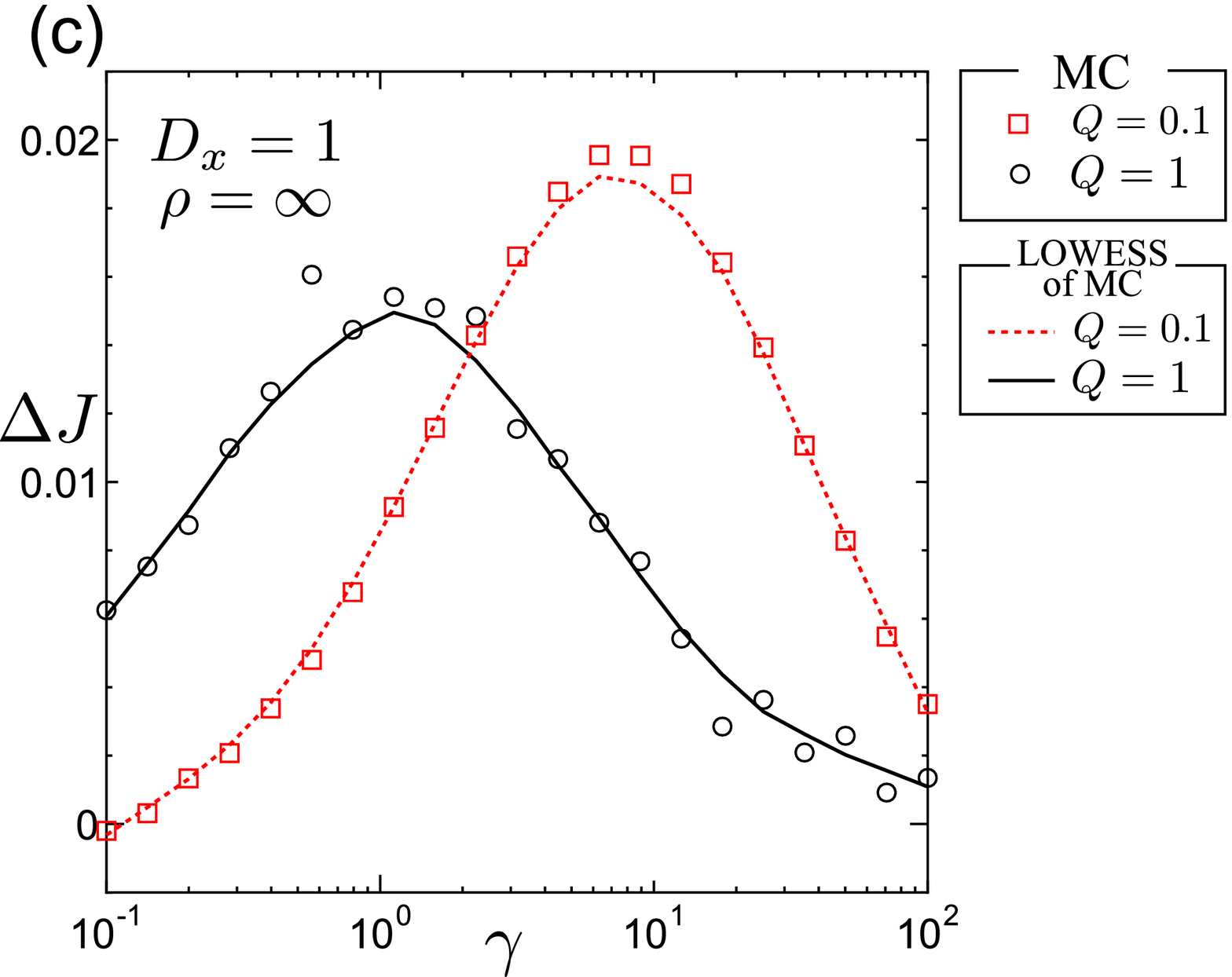} 
\par\end{centering}

\caption{(Online version in color) (a) and (b) Current $J$ with load $F>0$
as a function of $\gamma$ using MC simulations for (a) $D_{x}=1$,
$Q=0.1$, $\rho=\infty$, and $F=0.025$; and (b) $D_{x}=1$, $Q=1$,
$\rho=\infty$ and $F=0.006$. $J$ is calculated with MC simulations
with $\mu=0.01$ (circles) and $\mu=1$ (squares), and the overdamped
case ($\mu=0$) is also shown (solid line). (c) $\Delta J$ (difference
of the currents for $\mu=0.01$ and $1$ defined by Equation~\ref{eq:DeltaJ_def})
as a function $\gamma$ for two mass separations. Lines represent
LOWESS (locally weighted scatterplot smoothing) of MC simulations
(see the text). \label{fig:mass_sep}}
\end{figure}

We next show the dependence of the average current $J$ on $\gamma$
for two values of $Q$ (the effective noise intensity): Case (a) $Q=0.1$
{[}$D_{x}=1$, $\rho=\infty$, and $F=0.025${]} (Figure~\ref{fig:mass_sep}(a))
and Case (b) $Q=1$ {[}$D_{x}=1$, $\rho=\infty$, and $F=0.006${]}
(Figure~\ref{fig:mass_sep}(b)). In both cases, the load $F$ {[}Case
(a) $F=0.025$ and Case (b) $F=0.006${]} increased the ratchet potential
for larger values of $x$ and the value of $F$ was determined such
that particles of $\mu=0.01$ and $\mu=1$ move in opposite directions.
Figure~\ref{fig:mass_sep} shows $J$ of $\mu=0.01$ (circles) and
$\mu=1$ (squares) for $Q=0.1$ (Figure~\ref{fig:mass_sep}(a)) and
$Q=1$ (Figure~\ref{fig:mass_sep}(b)), where the lines show the
MCFM results for the overdamped case ($\mu=0$). We include the $\mu=0.01$
case along with the MCFM result to show that results of the underdamped
case decrease asymptotically to the overdamped case as $\mu\rightarrow0$.
In Cases (a) and (b), the current $J$ for $\mu=0.01$ has positive
values, unlike $J$ for $\mu=1$, which is negative over the whole
range of $\gamma$. $J$ for $\mu=0.01$ has peaks at intermediate
$\gamma$ values; however, the $\gamma_{\max}$ values at which the
current is a maximum are different in Cases (a) and (b): $\gamma_{\max}$
in Case (a) ($Q=0.1$) is around $10$, whereas $\gamma_{\max}$ in
Case (b) ($Q=1$) is about $1$. Since the currents for $\mu=0.01$
and $\mu=1$ move in the opposite directions, particles with $\mu=0.01$
and $\mu=1$ can be set apart due to the mass separation effect. To
quantify the mass separation capability, we define $\Delta J$ as
\begin{equation}
\Delta J=J(\mu=0.01)-J(\mu=1).\label{eq:DeltaJ_def}
\end{equation}
$\Delta J$ is a quantity of interest for the separation phenomenon,
and systems with larger $\Delta J$ exhibit a better separation capability.
Figure~\ref{fig:mass_sep}(c) shows $\Delta J$ for two cases {[}Case
(a) ($Q=0.1$) and Case (b) ($Q=1$){]} and the lines show locally
weighted scatterplot smoothing (LOWESS) for each data. It clearly
shows that $\gamma$ at which the mass separation capability is a
maximum is different for the two $Q$ cases and that better mass separation
is realized at smaller $\gamma$ values in the larger $Q$ case. This
result shows that the mass separation capability is greater for slower
environmental fluctuations when Brownian motors are subject to strong
noise.

\section{Discussion}

We investigated the statistical properties of SIN on Brownian ratchet
and found that the current was enhanced for smaller $\gamma$ in both
the overdamped and underdamped regimes (Figures~\ref{fig:OD_v_Q}
and \ref{fig:mass_sep}) when the effective noise-intensity $Q$ is
larger. This result is intriguing because a smaller $\gamma$ corresponds
to extrinsic fluctuations with a larger time scale. In single-cell
experiments on \emph{E. coli}, assuming that mRNA production is Poissonian
and that the protein burst size has an exponential distribution, the
protein copy number obeys the gamma distribution \cite{Taniguchi:2010:SingleCell}:
\begin{equation}
P(x)=\frac{x^{a-1}e^{-x/b}}{\Gamma(a)b^{a}},\label{eq:Gamma_def}
\end{equation}
where $a$ and $b$ are parameters. Observations \cite{Taniguchi:2010:SingleCell}
revealed that the fluctuations (extrinsic noise) in $a$ and $b$
are slow and the stationary distribution can be approximated as 
\begin{equation}
P(x)=\int_{0}^{\infty}db\int_{0}^{\infty}da\,\frac{x^{a-1}e^{-x/b}}{\Gamma(a)b^{a}}P(a)P(b),\label{eq:SS_Poisson}
\end{equation}
where $P(a)$ and $P(b)$ denote the distributions of $a$ and $b$,
respectively. Note that Equation~\ref{eq:SS_Poisson} is equivalent
to the description of superstatistics (see Equation~\ref{eq:superstat_def}).
In stochastic gene expression, the intrinsic noise is rather well-explained:
its source is stochastic chemical reactions with small number of molecules.
In contrast, contribution of the extrinsic noise has not been well
reasoned or modeled \cite{Scott:2006:ExtrinsicNoise,Shahrezaei:2008:ColoredGene}.
Stochastic gene expression is often modeled with Langevin equations,
i.e., continuous approximation of the continuous-time Markov chain
through Van Kampen's expansion~\cite{Kampen:1992:StocProcBook}.
Many models assume that extrinsic noise affect the transcriptional,
translational and degradation kinetics as in Equation~\ref{eq:SS_Poisson},
which results in both drift terms and noise-intensity fluctuations
of corresponding Langevin equations. The stochastic term of Langevin
equations is often approximated with the additive white noise, and
therefore, the extrinsic noise-induced fluctuation fits the overdamped
case given by Equation~\ref{eq:UD_Langevin_x} and \ref{eq:UD_Langevin_s},
assuming that the drift term fluctuation is negligible compared to
that of the noise-intensity. Our calculations showed that environmental
fluctuations should be slow to enhance the transport capability in
noisy environments. That is, fluctuations in $a$ and $b$ in Equation~\ref{eq:Gamma_def}
occur over long time scales.

Lastly, let us show the correspondence of our calculations with actual
time-scales in biological experiments. Refs.~\cite{Rosenfeld:2005:SingleCell}
observed that the time correlation of extrinsic fluctuations in \emph{E.
coli} is in the order of cell cycle length $T_{cc}$. The deterministic
part of the protein concentration $x$ generally obeys 
\begin{equation}
\frac{dx}{dt}=f_{+}-\beta x,\label{eq:linear_model}
\end{equation}
where $f_{+}$ is a protein synthesis term (via translation) and $\beta$
is the degradation rate. In Langevin equations, the degradation rate
corresponds to the relaxation rate (reciprocal of relaxation time)
and hence it determines the time scale. Dunlop \textit{et al.}~\cite{Dunlop:2008:GeneNoiseCor}
defined $\beta=\log2/T_{cc}$ and concluded that the time correlation
of the extrinsic fluctuation is $\tau=T_{cc}/\log2$ ($T_{cc}=60\min$).
Because the ratchet potential given by Equation~\ref{eq:potential}
can be approximated by a quadratic function around the minima, Equation~\ref{eq:UD_Langevin_x}
can be cast in the form of Equation~\ref{eq:linear_model} with $\beta\simeq10$.
Matching the time scales of these two systems by comparing the relaxation
time (i.e., comparing $\beta$ in the two cases), the time correlation
of the extrinsic fluctuations observed in \emph{E. coli} corresponds
to $\tau=0.1$ in our ratchet potential model. Figure~\ref{fig:OD_v_Q}
shows that $\gamma_{\max}$ is in the approximate range of $1$--$10$,
which implies that $\tau_{\max}=\gamma_{\max}^{-1}=0.1\sim1$. Therefore,
the extrinsic noise in the experimental observation is close to optimal
for the transport effect. Extrinsic noise has non-negligible time-correlation,
leaving the system at nonequilibrium states. Although there remains
a biochemical gap between gene expressions and ion transports, our
result shows the biochemical advantage of exploiting extrinsic noise
for gene regulation.

In summary, we investigated the transport properties of Brownian ratchet
in both overdamped and underdamped regimes~\cite{Hasegawa:2010:SIN,Hasegawa:2011:BistableSIN}.
In the overdamped regime, our calculations by the MCFM and MC simulations
revealed the existence of a maximum current as a function of $\gamma$
(the relaxation rate) and $Q$ (the effective noise intensity). The
maximum current is induced at a lower relaxation rate $\gamma$ for
higher noise intensities. In the underdamped regime, MC calculations
also showed a maximum for smaller $\gamma$ when systems are subject
to noisy environments. Consequently, the mass separation capability
was also maximized for smaller $\gamma$ in such cases. We continue
the investigation of ratchet transport in specific biological models
in future studies.

\section*{Acknowledgments}

This work was supported by a Global COE program ``Deciphering Biosphere
from Genome Big Bang'' from the Ministry of Education, Culture, Sports,
Science and Technology (MEXT), Japan (YH) and a Grant-in-Aid for Young
Scientists B (\#23700263) from MEXT, Japan (YH).

\appendix

\section{Kurtosis of SIN\label{sec:kurtosis}}

We calculate kurtosis $\kappa$ of SIN defined by 
\begin{equation}
\kappa=\frac{\left\langle \left\{ s(t)\xi_{x}(t)\right\} ^{4}\right\rangle }{\left\langle \left\{ s(t)\xi_{x}(t)\right\} ^{2}\right\rangle ^{2}}.\label{eq:kappa_def}
\end{equation}
By using an independence relation between $s$ and $\xi_{x}$ and
the Gaussian nature of $\xi_{x}$, the numerator and denominator in
Equation~\ref{eq:kappa_def} are given by 
\begin{eqnarray}
\left\langle \left\{ s(t)\xi_{x}(t)\right\} ^{4}\right\rangle  & = & 3\left\langle s(t)^{4}\right\rangle \left\langle \xi_{x}(t)^{2}\right\rangle ^{2},\label{eq:numerator}\\
\left\langle \left\{ s(t)\xi_{x}(t)\right\} ^{2}\right\rangle ^{2} & = & \left\langle s(t)^{2}\right\rangle ^{2}\left\langle \xi_{x}(t)^{2}\right\rangle ^{2}.\label{eq:denominator}
\end{eqnarray}
Let $u(t)=s(t)-\alpha$. Since $u(t)$ is a standard Ornstein--Uhlenbeck
process with $\left\langle u(t)\right\rangle =0$, the second and
fourth-order moments are $\left\langle u(t)^{2}\right\rangle =D_{s}$
and $\left\langle u(t)^{4}\right\rangle =3D_{s}^{2}$, which yields
\begin{eqnarray}
\left\langle s(t)^{2}\right\rangle  & = & \left\langle u(t)^{2}\right\rangle +\alpha^{2}=D_{s}+\alpha^{2},\label{eq:OU_second_moment}\\
\left\langle s(t)^{4}\right\rangle  & = & \alpha^{4}+6\alpha^{2}\left\langle u(t)^{2}\right\rangle +\left\langle u(t)^{4}\right\rangle =\alpha^{4}+6\alpha^{2}D_{s}+3D_{s}^{2}.\label{eq:OU_fourth_moment}
\end{eqnarray}
With Equations~\ref{eq:numerator}--\ref{eq:OU_fourth_moment}, the
kurtosis is calculated into 
\begin{equation}
\kappa=\frac{3\left(\alpha^{4}+6\alpha^{2}D_{s}+3D_{s}^{2}\right)}{\left(D_{s}+\alpha^{2}\right)^{2}}.\label{eq:kurtosis}
\end{equation}
Substituting $\rho=D_{s}/\alpha^{2}$ into Equation~\ref{eq:kurtosis},
we obtain Equation~\ref{eq:kurtosis_rho}.

\section{Matrix continued fraction method\label{sec:MCFM}}

We explain the procedure of the MCFM in the overdamped case. Substituting
Equation~\ref{eq:statdist_overdamped} into Equation~\ref{eq:FPE_OD},
we obtain a linear algebraic equation: 
\begin{eqnarray}
0 & = & C_{k,n}\left[-n\gamma+2F\pi k\mathrm{i}-4\pi^{2}D_{x}k^{2}\left\{ \alpha^{2}+2D_{s}\left(n+\frac{1}{2}\right)\right\} \right]\nonumber \\
 &  & +C_{k-1,n}\pi k\mathrm{i}+C_{k+1,n}\pi k\mathrm{i}+\frac{C_{k-2,n}}{2}\pi k\mathrm{i}+\frac{C_{k+2,n}}{2}\pi k\mathrm{i}\nonumber \\
 &  & -4C_{k,n+2}\pi^{2}D_{x}D_{s}k^{2}\sqrt{(n+2)(n+1)}-4C_{k,n-2}\pi^{2}D_{x}D_{s}k^{2}\sqrt{n(n-1)}\nonumber \\
 &  & -8C_{k,n-1}\pi^{2}D_{x}\alpha k^{2}\sqrt{D_{s}n}-8C_{k,n+1}\pi^{2}D_{x}\alpha k^{2}\sqrt{D_{s}(n+1)},\label{eq:linalg_eq}
\end{eqnarray}
where the dimension of the linear algebraic equation (\ref{eq:linalg_eq})
is $(2M_{k}+1)(M_{n}+1)$. We show the procedure of the MCFM, since
a naive conversion of Equation~\ref{eq:linalg_eq} yields an equation
with second-nearest-neighbor coupling, which is not compatible with
the MCFM (the MCFM can treat only first-nearest-neighbor coupling).
By introducing $c_{k}=(C_{k,0},C_{k,1}\cdots,C_{k,M_{n}})^{\mathsf{T}}$,
Equation~\ref{eq:linalg_eq} is calculated into 
\begin{equation}
0=\frac{\pi k\mathrm{i}\mathsf{E}c_{k-2}}{2}+\pi k\mathrm{i}\mathsf{E}c_{k-1}+\left[\mathsf{A}-4\pi^{2}D_{x}k^{2}\mathsf{B}+2F\pi k\mathrm{i}\mathsf{E}\right]c_{k}+\pi k\mathrm{i}\mathsf{E}c_{k+1}+\frac{\pi k\mathrm{i}\mathsf{E}c_{k+2}}{2}.\label{eq:recurrence_rel}
\end{equation}
Here, $\mathsf{E}$ is the identity matrix and $\mathsf{A}$ and $\mathsf{B}$
are $(M_{n}+1)\times(M_{n}+1)$ matrices defined by 
\begin{eqnarray}
\mathsf{A}_{n+1,n^{\prime}+1} & = & -n\gamma\delta_{n,n^{\prime}},\label{eq:A_mat}\\
\mathsf{B}_{n+1,n^{\prime}+1} & = & \left\{ \alpha^{2}+2D_{s}\left(n+\frac{1}{2}\right)\right\} \delta_{n,n^{\prime}}+2\alpha\sqrt{D_{s}n}\delta_{n-1,n^{\prime}}+2\alpha\sqrt{D_{s}(n+1)}\delta_{n+1,n^{\prime}},\nonumber \\
 &  & +D_{s}\sqrt{n(n-1)}\delta_{n-2,n^{\prime}}+D_{s}\sqrt{(n+1)(n+2)}\delta_{n+2,n^{\prime}},\label{eq:B_mat}
\end{eqnarray}
where $0\le n\le M_{n}$ and $0\le n^{\prime}\le M_{n}$. Introducing
$\tilde{c}_{k}=(c_{2k}^{\mathsf{T}},c_{2k+1}^{\mathsf{T}})^{\mathsf{T}}$
\cite{Risken:1989:FPEBook}, Equation~\ref{eq:recurrence_rel}, which
is an equation with second-nearest-neighbor coupling, reduces to the
following equation with first-nearest-neighbor coupling: 
\begin{equation}
0=\mathsf{Q}_{k}^{-}\tilde{c}_{k-1}+\mathsf{Q}_{k}\tilde{c}_{k}+\mathsf{Q}_{k}^{+}\tilde{c}_{k+1},\label{eq:1_recur}
\end{equation}
where $\mathsf{Q}_{k}$ are $(2M_{n}+2)\times(2M_{n}+2)$ matrices
consisting of submatrices $\mathsf{A}$ and $\mathsf{B}$: 
\begin{eqnarray}
\mathsf{Q}_{k}^{-} & = & \left(\begin{array}{cc}
\pi k\mathrm{i}\mathsf{E} & 2\pi k\mathrm{i}\mathsf{E}\\
\mathsf{0} & \pi(2k+1)\mathrm{i}\mathsf{E}/2
\end{array}\right),\label{eq:Qn_mat}\\
\mathsf{Q}_{k} & = & \left(\begin{array}{cc}
\mathsf{A}-16\pi^{2}D_{x}k^{2}\mathsf{B}+4F\pi k\mathrm{i}\mathsf{E} & 2\pi k\mathrm{i}\mathsf{E}\\
\pi(2k+1)\mathrm{i}\mathsf{E} & \mathsf{A}-4\pi^{2}D_{x}(2k+1)^{2}\mathsf{B}+2F\pi(2k+1)\mathrm{i}\mathsf{E}
\end{array}\right),\label{eq:Q_mat}\\
\mathsf{Q}_{k}^{+} & = & \left(\begin{array}{cc}
\pi k\mathrm{i}\mathsf{E} & \mathsf{0}\\
\pi(2k+1)\mathrm{i}\mathsf{E} & \pi(2k+1)\mathrm{i}\mathsf{E}/2
\end{array}\right).\label{eq:Qp_mat}
\end{eqnarray}
We solve the recurrence relation for Equation~\ref{eq:1_recur} by
introducing $\mathsf{S}_{k}$ and $\mathsf{R}_{k}$ that satisfy $\tilde{c}_{k+1}=\mathsf{S}_{k}\tilde{c}_{k}$
($k\ge0$) and $\tilde{c}_{k-1}=\mathsf{R}_{k-1}\tilde{c}_{k}$ ($k\le0$).
With $\mathsf{S}_{k}$ and $\mathsf{R}_{k}$, Equation~\ref{eq:1_recur}
is calculated into
\begin{eqnarray}
\mathsf{S}_{k-1} & = & -\left(\mathsf{Q}_{k}+\mathsf{Q}_{k}^{+}\mathsf{S}_{k}\right)^{-1}\mathsf{Q}_{k}^{-}\hspace{1em}(k\ge1),\label{eq:S_rel}\\
\mathsf{R}_{k} & = & -\left(\mathsf{Q}_{k}^{-}\mathsf{R}_{k-1}+\mathsf{Q}_{k}\right)^{-1}\mathsf{Q}_{k}^{+}\hspace{1em}(k\le-1),\label{eq:R_rel}
\end{eqnarray}
where $\mathsf{S}_{k}$ and $\mathsf{R}_{k}$ can be obtained by truncating
at large $k$, namely at $k=-\tilde{M}_{k},\tilde{M}_{k}(\approx M_{k}/2)$.
For $k=0$, we have 
\begin{equation}
0=\left[\mathsf{Q}_{0}^{-}\mathsf{R}_{-1}+\mathsf{Q}_{0}+\mathsf{Q}_{0}^{+}\mathsf{S}_{0}\right]\tilde{c}_{0},\label{eq:neib_k0}
\end{equation}
where the first row of the left part of matrix in Equation~\ref{eq:neib_k0}
vanishes due to Equations~\ref{eq:Qn_mat}--\ref{eq:Qp_mat}. Therefore,
$\tilde{c}_{0}$ has a nontrivial solution and $\tilde{c}_{k}$ can
be calculated by recursively applying $\mathsf{S}_{k}$ and $\mathsf{R}_{k}$
to $\tilde{c}_{0}$.


\begin{thebibliography}{10}

\bibitem{Benzi:1981:SR}
R.~Benzi, A.~Sutera, and A.~Vulpiani.
\newblock The mechanism of stochastic resonance.
\newblock {\em J. Phys. A}, 14:L453, 1981.

\bibitem{McNamara:1989:SR}
B.~McNamara and K.~Wiesenfeld.
\newblock Theory of stochastic resonance.
\newblock {\em Phys. Rev. A}, 39:4854--4869, 1989.

\bibitem{Jung:1991:AmpSR}
P.~Jung and P.~H\"anggi.
\newblock Amplification of small signals via stochastic resonance.
\newblock {\em Phys. Rev. A}, 44:8032--8042, 1991.

\bibitem{Gammaitoni:1998:SR}
L.~Gammaitoni, P.~H\"anggi, P.~Jung, and F.~Marchesoni.
\newblock Stochastic resonance.
\newblock {\em Rev. Mod. Phys.}, 70:223--287, 1998.

\bibitem{McDonnell:2008:SRBook}
M.~D. McDonnell, N.~G. Stocks, C.~E.~M. Pearce, and D.~Abbott.
\newblock {\em Stochastic resonance}.
\newblock Cambridge University Press, 2008.

\bibitem{McDonnell:2009:SR}
M.~D. McDonnell and D.~Abbott.
\newblock What is stochastic resonance? definitions, misconceptions, debates,
  and its relevance to biology.
\newblock {\em PLoS Comput. Biol.}, 5:e1000348, 2009.

\bibitem{Marchesoni:1996:SpatiotempSR}
F.~Marchesoni, L.~Gammaitoni, and A.~R. Bulsara.
\newblock Spatiotemporal stochastic resonance in a $\phi^4$ model of
  kink-antikink nucleation.
\newblock {\em Phys. Rev. Lett.}, 76:2609--2612, 1996.

\bibitem{Teramae:2004:NoiseIndSync}
J.~Teramae and D.~Tanaka.
\newblock Robustness of the noise-induced phase synchronization in a general
  class of limit cycle oscillators.
\newblock {\em Phys. Rev. Lett.}, 93:204103, 2004.

\bibitem{Acebron:2005:KuramotoReview}
J.~A. Acebr\'on, L.~L. Bonilla, C.~J. P\'erez~Vicente, F.~Ritort, and
  R.~Spigler.
\newblock The {Kuramoto} model: A simple paradigm for synchronization
  phenomena.
\newblock {\em Rev. Mod. Phys.}, 77:137--185, Apr 2005.

\bibitem{Nakao:2007:NISinLC}
H.~Nakao, K.~Arai, and Y.~Kawamura.
\newblock Noise-induced synchronization and clustering in ensembles of
  uncoupled limit-cycle oscillators.
\newblock {\em Phys. Rev. Lett.}, 98:184101, 2007.

\bibitem{Koern:2005:GeneNoiseReview}
M.~K{\oe}rn, T.~C. Elston, W.~J. Blake, and J.~J. Collins.
\newblock Stochasticity in gene expression: from theories to phenotypes.
\newblock {\em Nat. Rev.}, 6:451--464, 2005.

\bibitem{Patnaik:2006:NoiseReview}
P.~R. Patnaik.
\newblock External, extrinsic and intrinsic noise in cellular systems:
  analogies and implications for protein synthesis.
\newblock {\em Biotechnol. Mol. Biol. Rev.}, 1:121--127, 2006.

\bibitem{Maheshri:2007:LivingNoisyGene}
N.~Maheshri and E.~K. O'Shea.
\newblock Living with noisy genes: How cells function reliably with inherent
  variability in gene expression.
\newblock {\em Annu. Rev. Biophys. Biomol. Struct.}, 36:413--434, 2007.

\bibitem{Rausenberger:2009:GeneNoiseReview}
J.~Rausenberger, C.~Fleck, J.~Timmer, and M.~Kollmann.
\newblock Signatures of gene expression noise in cellular systems.
\newblock {\em Prog. Biophys. Mol. Biol.}, 100:57--66, 2009.

\bibitem{Taniguchi:2010:SingleCell}
Y.~Taniguchi, P.~J. Choi, G.-W. Li, H.~Chen, M.~Babu, J.~Hearn, A.~Emili, and
  X.~S. Xie.
\newblock Quantifying \emph{{E}. coli} proteome and transcriptome with
  single-molecule sensitivity in single cells.
\newblock {\em Science}, 329:533, 2010.

\bibitem{Jeffrey:2007:CyanoBac}
J.~R. Chabot, J.~M. Pedraza, P.~Luitel, and A.~van Oudenaarden.
\newblock Stochastic gene expression out-of-steady-state in the cyanobacterial
  circadian clock.
\newblock {\em Nature}, 450:1249--1252, 2007.

\bibitem{Nozaki:1999:ColoredNoiseNeuron}
D.~Nozaki, D.~J. Mar, P.~Grigg, and J.~J. Collins.
\newblock Effects of colored noise on stochastic resonance in sensory neurons.
\newblock {\em Phys. Rev. Lett.}, 82:2402--2405, Mar 1999.

\bibitem{Fuentes:2001:nonGaussSR}
M.~A. Fuentes, R.~Toral, and H.~S. Wio.
\newblock Enhancement of stochastic resonance: the role of non {Gaussian}
  noises.
\newblock {\em Physica A}, 295:114--122, 2001.

\bibitem{Magnasco:1993:ThermalRatchet}
M.~O. Magnasco.
\newblock Forced thermal ratchets.
\newblock {\em Phys. Rev. Lett.}, 71:1477--1481, Sep 1993.

\bibitem{Hanggi:2009:BrownianMotorsReview}
P.~H\"anggi and F.~Marchesoni.
\newblock Artificial {Brownian} motors: Controlling transport on the nanoscale.
\newblock {\em Rev. Mod. Phys.}, 81:387--442, 2009.

\bibitem{Astumian:1994:MolecularMotor}
R.~D. Astumian and M.~Bier.
\newblock Fluctuation driven ratchets: molecular motors.
\newblock {\em Phys. Rev. Lett.}, 72:1766--1769, 1994.

\bibitem{Astumian:1997:BrownMotor}
R.~D. Astumian.
\newblock Thermodynamics and kinetics of a {Brownian} motor.
\newblock {\em Science}, 276:917--922, 1997.

\bibitem{Astumian:1998:MotorAndPump}
R.~D. Astumian and I.~Der\'enyi.
\newblock Fluctuation driven transport and models of molecular motors and
  pumps.
\newblock {\em Eur. Biophys. J}, 27:474--489, 1998.

\bibitem{Tsong:2002:NaK_ATPase}
T.~Y. Tsong.
\newblock Na, {K-ATPase} as a {Brownian} motor: Electirc field-induced
  conformational fluctuation leads to uphill pumping of cation in the absence
  of {ATP}.
\newblock {\em J. Biol. Phys.}, 28:309--325, 2002.

\bibitem{Marchesoni:1998:MassSep}
F.~Marchesoni.
\newblock Conceptual design of a molecular shuttle.
\newblock {\em Phys. Lett. A}, 237:126--130, 1998.

\bibitem{Linke:2002:QuantumRatchet}
H.~Linke, T.~E. Humphrey, P.~E. Lindelof, A.~Lofgren, R.~Newbury, P.~Omling,
  A.~O. Sushkov, R.~P. Taylor, and H.~Xu.
\newblock Quantum ratchets and quantum heat pumps.
\newblock {\em Appl. Phys. A}, 75:237--246, 2002.

\bibitem{Lundh:2005:RatchetOptLat}
E.~Lundh and M.~Wallin.
\newblock Ratchet effect for cold atoms in an optical lattice.
\newblock {\em Phys. Rev. Lett.}, 94:110603, 2005.

\bibitem{Jose:2005:RandomWalker}
J.~L. Mateos.
\newblock A random walker on a ratchet.
\newblock {\em Physica A}, 351:79--87, 2005.

\bibitem{Bouzat:2004:nonGaussMotor}
S.~Bouzat and H.~S. Wio.
\newblock Current and efficiency enhancement in {Brownian} motors driven by non
  {Gaussian} noises.
\newblock {\em Eur. Phys. J. B}, 41:97--105, 2004.

\bibitem{Mangioni:2008:WalkerRatchetNGN}
S.~E. Mangioni and H.~S. Wio.
\newblock A random walker on a ratchet potential: effect of a non {Gaussian}
  noise.
\newblock {\em Eur. Phys. J. B}, 61:67--73, 2008.

\bibitem{Frey:2005:BrownianRev}
E.~Frey and K.~Kroy.
\newblock {Brownian} motion: a paradigm of soft matter and biological physics.
\newblock {\em Ann. Phys.}, 14:20--50, 2005.

\bibitem{Wilk:2000:NEXTParam}
G.~Wilk and Z.~W\l{}odarczyk.
\newblock Interpretation of the nonextensivity parameter $q$ in some
  applications of {Tsallis} statistics and {L}\'evy distributions.
\newblock {\em Phys. Rev. Lett.}, 84:2770--2773, Mar 2000.

\bibitem{Beck:2001:DynamicalNEXT}
C.~Beck.
\newblock Dynamical foundations of nonextensive statistical mechanics.
\newblock {\em Phys. Rev. Lett.}, 87:180601, Oct 2001.

\bibitem{Beck:2003:Superstatistics}
C.~Beck and E.~G.~D. Cohen.
\newblock Superstatistics.
\newblock {\em Physica A}, 322:267--275, 2003.

\bibitem{Beck:2011:Superstatistics}
C.~Beck.
\newblock Generalized statistical mechanics for superstatistical systems.
\newblock {\em Phil. Trans. R. Soc. A}, 369:453--465, 2011.

\bibitem{Tsallis:1988:Generalization}
C.~Tsallis.
\newblock Possible generalization of {Boltzmann}--{Gibbs} statistics.
\newblock {\em J. Stat. Phys.}, 52:479--487, 1988.

\bibitem{Tsallis:2009:NonextensiveBook}
C.~Tsallis.
\newblock {\em Introduction to Nonextensive Statistical Mechanics: Approaching
  a Complex World}.
\newblock Springer, 2009.

\bibitem{Queiros:2005:VolatilityNEXT}
S.~M.~D. Queir\'os and C.~Tsallis.
\newblock On the connection between financial processes with stochastic
  volatility and nonextensive statistical mechanics.
\newblock {\em Eur. Phys. J. B}, 48:139--148, 2005.

\bibitem{Beck:2006:SS_Brownian}
C.~Beck.
\newblock Superstatistical {Brownian} motion.
\newblock {\em Prog. Theor. Phys. Suppl.}, 162:29--36, 2006.

\bibitem{Rodriguez:2007:SS_Brownian}
R.~F. Rodr\'iguez and I.~Santamar\'ia-Holek.
\newblock Superstatistics of {Brownian} motion: A comparative study.
\newblock {\em Physica A}, 385:456--464, 2007.

\bibitem{Jizba:2008:SuposPD}
P.~Jizba and H.~Kleinert.
\newblock Superpositions of probability distributions.
\newblock {\em Phys. Rev. E}, 78:031122, 2008.

\bibitem{Hasegawa:2010:qExpBistable}
Y.~Hasegawa and M.~Arita.
\newblock Bistable stochastic processes in the $q$-exponential family.
\newblock {\em Physica A}, 389:4450--4461, 2010.

\bibitem{Yalcin:2012:PolymerSS}
G.~C. Yalcin and C.~Beck.
\newblock Currents in complex polymers: An example of superstatistics for short
  time series.
\newblock {\em Phys. Lett. A}, 376:2344--2347, 2012.

\bibitem{Hasegawa:2010:SIN}
Y.~Hasegawa and M.~Arita.
\newblock Noise-intensity fluctuation in {Langevin} model and its higher-order
  {Fokker}--{Planck} equation.
\newblock {\em Physica A}, 390:1051--1063, 2011.

\bibitem{Hasegawa:2011:BistableSIN}
Y.~Hasegawa and M.~Arita.
\newblock Escape process and stochastic resonance under noise-intensity
  fluctuation.
\newblock {\em Phys. Lett. A}, 375:3450--3458, 2011.

\bibitem{Reimann:1996:OscBrownianMot}
P.~Reimann, R.~Bartussek, R.~H{\"a}u{\ss}ler, and P.~H{\"a}nggi.
\newblock Brownian motors driven by temperature oscillations.
\newblock {\em Phys. Lett. A}, 215:26--31, 1996.

\bibitem{Li:1997:Transport}
Y.-X. Li.
\newblock Transport generated by fluctuating temperature.
\newblock {\em Physica A}, 238:245--251, 1997.

\bibitem{Zhang:2008:OscTempRatchet}
Y.~Zhang and J.~Chen.
\newblock Investigation on a temporal asymmetric oscillating temperature
  ratchet.
\newblock {\em Physica A}, 387:3443--3448, 2008.

\bibitem{Borromeo:2006:AutoMaxDeamon}
M.~Borromeo, S.~Giusepponi, and F.~Marchesoni.
\newblock Recycled noise rectification: An automated {Maxwell}'s deamon.
\newblock {\em Phys. Rev. E}, 74:031121, 2006.

\bibitem{Morgado:2011:MassPoisson}
W.~A.~M. Morgado, S.~M.~D. Queir\'os, and D.~O. Soares-Pinto.
\newblock On exact time averages of a massive {Poisson} particle.
\newblock {\em J. Stat. Mech.}, page P06010, 2011.

\bibitem{Doering:1992:ResonantActivation}
C.~R. Doering and J.~C. Gadoua.
\newblock Resonant activation over a fluctuating barrier.
\newblock {\em Phys. Rev. Lett.}, 69:2318--2321, Oct 1992.

\bibitem{Mantegna:1996:NES}
R.~N. Mantegna and B.~Spagnolo.
\newblock Noise enhanced stability in an unstable system.
\newblock {\em Phys. Rev. Lett.}, 76:563--566, Jan 1996.

\bibitem{Spagnolo:2008:NES_review}
B.~Spagnolo, N.~V. Agudov, and A.~A. Dubkov.
\newblock Noise enhanced stability.
\newblock {\em Acta Phys. Pol. B}, 35:1419--1436, 2004.

\bibitem{Rosenfeld:2005:SingleCell}
N.~Rosenfeld, J.~W. Young, U.~Alon, P.~S. Swain, and M.~B. Elowitz.
\newblock Gene regulation at the single-cell level.
\newblock {\em Science}, 307:1962--1965, 2005.

\bibitem{Dunlop:2008:GeneNoiseCor}
M.~J. Dunlop, R.~S.~C. III, J.~H. Levine, R.~M. Murray, and M.~B. Elowitz.
\newblock Regulatory activity revealed by dynamic correlations in gene
  expression noise.
\newblock {\em Nature Genetics}, 40:14393--1498, December 2008.

\bibitem{Bartussek:1996:CorRatchet}
R.~Bartussek, P.~Reimann, and P.~H\"anggi.
\newblock Precise numerics versus theory for correlation ratchets.
\newblock {\em Phys. Rev. Lett.}, 76:1166--1169, 1996.

\bibitem{Lindner:1999:InertiaRatchet}
B.~Lindner, L.~Schimansky-Geier, P.~Reimann, P.~H\"anggi, and M.~Nagaoka.
\newblock Inertia ratchets: A numerical study versus theory.
\newblock {\em Phys. Rev. E}, 59:1417--1424, 1999.

\bibitem{Risken:1989:FPEBook}
H.~Risken.
\newblock {\em The {Fokker}--{Planck} Equation: Methods of Solution and
  Applications}.
\newblock Springer, 2nd edition, 1989.

\bibitem{Scott:2006:ExtrinsicNoise}
M.~Scott, B.~Ingalls, and M.~K{\ae}rn.
\newblock Estimations of intrinsic and extrinsic noise in models of nonlinear
  genetic networks.
\newblock {\em Chaos}, 16:026107, 2006.

\bibitem{Shahrezaei:2008:ColoredGene}
V.~Shahrezaei, J.~F. Ollivier, and P.~S. Swain.
\newblock Colored extrinsic fluctuations and stochastic gene expression.
\newblock {\em Mol. Syst. Biol.}, 4:196, 2008.

\bibitem{Kampen:1992:StocProcBook}
V.~Kampen.
\newblock {\em Stochastic Process Theory in Physics and Chemistry}.
\newblock North-Holland, 1992.

\end{thebibliography}
\end{document}